\documentclass[apjl,iop,twocolappendix]{emulateapj}

\usepackage{threeparttable}

\usepackage[table,xcdraw]{xcolor}
\usepackage{diagbox}
\usepackage{amssymb}
\usepackage{mathtools}
\usepackage{color}
\usepackage{xifthen}
\newcommand{\toto}[1][]{ %
\ifthenelse{\isempty{#1}}{{\color{red} Missing}}{{\color{red} #1}} %
}

\usepackage[]{bm}

\shorttitle{ACAD-OSM I}
\shortauthors{J. Mazoyer et al.}
\usepackage[colorlinks,breaklinks=false, citecolor = {blue}, urlcolor = {blue}]{hyperref}

\begin{document}

\title{Active correction of aperture discontinuities - optimized stroke minimization I: \\a new adaptive interaction matrix algorithm}

\author{J.~Mazoyer\altaffilmark{1,2}, L.~Pueyo\altaffilmark{2}, M.~N'Diaye\altaffilmark{3}, K.~Fogarty\altaffilmark{1}, N.~Zimmerman\altaffilmark{2,4}, L.~Leboulleux\altaffilmark{2,5,6}, K. E.~St. Laurent\altaffilmark{2}, R.~Soummer\altaffilmark{2}, S.~Shaklan\altaffilmark{7} and C.Norman\altaffilmark{1}}
\altaffiltext{1}{Johns Hopkins University, Zanvyl Krieger School of Arts and Sciences, Department of Physics and Astronomy, Bloomberg Center for Physics and Astronomy, 3400 North Charles Street, Baltimore, MD 21218, USA}
\altaffiltext{2}{Space Telescope Science Institute, 3700 San Martin Dr, Baltimore MD 21218, USA}
\altaffiltext{3}{Universit\'e C\^ote d\'{}Azur, Observatoire de la C\^ote d\'{}Azur, CNRS, Laboratoire Lagrange, Bd de l\'{}Observatoire, CS 34229, 06304 Nice cedex 4, France}
\altaffiltext{4}{NASA Goddard Space Flight Center, Exoplanets and Stellar Astrophysics Laboratory, Greenbelt, MD 20771, USA}
\altaffiltext{5}{Aix Marseille Université, CNRS, LAM (Laboratoire d’Astrophysique de Marseille) UMR 7326, 13388, Marseille, France}
\altaffiltext{6}{Office National d\'{}Etudes et de Recherches A\'erospatiales, 29 Avenue de la Division Leclerc, 92320 Ch\^atillon, France}
\altaffiltext{7}{Jet Propulsion Laboratory, 4800 Oak Grove Drive, Pasadena, CA 91109, USA}

\email{jmazoyer@jhu.edu}

\begin{abstract}

{Future searches for biomarkers on habitable exoplanets will rely on telescope instruments that achieve extremely high contrast at small planet-to-star angular separations. Coronagraphy is a promising starlight suppression technique, providing excellent contrast and throughput for off-axis sources on clear apertures. However, the complexity of space- and ground-based telescope apertures goes on increasing over time, owing to the combination of primary mirror segmentation, secondary mirror, and support structures. These discontinuities in the telescope aperture limit the coronagraph performance.}
{In this paper, we present ACAD-OSM, a novel active method to correct for the diffractive effects of aperture discontinuities in the final image plane of a coronagraph. Active methods use one or several deformable mirrors that are controlled with an interaction matrix to correct for the aberrations in the pupil. However, they are often limited by the amount of aberrations introduced by aperture discontinuities. This algorithm relies on the recalibration of the interaction matrix during the correction process to overcome this limitation.}
{We first describe the ACAD-OSM technique and compare it to the previous active methods for the correction of aperture discontinuities. We then show its performance in terms of contrast and off-axis throughput for static aperture discontinuities (segmentation, struts) and for some aberrations evolving over the life of the instrument (residual phase aberrations, artifacts in the aperture, misalignments in the coronagraph design).}
{This technique can now obtain the earth-like planet detection threshold of $10^{−10}$ contrast on any given aperture over at least a 10\% spectral bandwidth, with several coronagraph designs.}

\end{abstract}


%-------------------------------------------------------------------------------------------------
%-------------------------------------------------------------------------------------------------
%-------------------------------------------------------------------------------------------------
\section{Introduction}
\label{sec:intro}
%-------------------------------------------------------------------------------------------------
%-------------------------------------------------------------------------------------------------
%-------------------------------------------------------------------------------------------------

The current generation of high-contrast imaging instruments \citep{macintosh08, beuzit08, martinache10,oppenheimer12} have already detected several exoplanets \citep{marois08, lagrange09, Kuzuhara13, rameau13b, bailey14, macintosh15} and obtained their first spectra \citep{bonnefoy16,chilcote17,rajan17}. Located on ground-based 8m class telescopes, these facilities benefit from important improvements in adaptive optics \citep[AO, ][]{sauvage07, wallace09} and coronagraphy \citep{soummer03,rouan00} in the past decade to reach $10^{-6}$ contrast levels and image young, bright Jovian planets. These companions orbiting nearby stars help us to better understand the formation and evolution of exo-planetary systems. The next generation of exoplanet imagers on the ground with Extremely Large Telescopes \citep[ELTs, ][]{macintosh06, kasper08,davies10,quanz15c} and in space with WFIRST \citep{spergel15} will seek $10^{-9}$ contrasts to enable the observation of older Jupiter-like planets, closing the gap with companions that are detected with other techniques. Further along, only envisioned space telescopes such as LUVOIR \citep{dalcanton15} or HabEx \citep{mennesson16} will aim for the $10^{-10}$ contrast required to detect Earth twins around nearby stars and investigate the presence of bio-markers in their atmosphere.

To reach these contrasts, several critical points must be addressed for both space and ground-based instruments. On the one hand, we need to correct for the quasi-static phase and amplitude aberrations. These errors are generated by the optics themselves and cannot be compensated for with classical AO system on ground-based telescopes if they are located in the science path after the beam splitter leading to the wavefront sensing channel. If uncorrected, these small aberrations limit the contrast to $10^{-6}$ in visible and near-IR optical systems using high quality optics. On the other hand, as the telescope primary mirror becomes larger, the complexity of the aperture geometry tends to increase. First, there is the central obscuration which can reach up to 36\% of the pupil diameter as in the case of WFIRST \citep{spergel15}. Second, as the telescope secondary mirror also becomes larger, its support structure becomes thicker, blocking more light and producing stronger diffractive effects in the star Point Spread Function (PSF). Finally, the segmentation of the aperture, already present on the Keck Observatory and the Gran Telescopio Canarias, will increase with up to hundreds of segments for the ELTs. Following JWST with 18 primary mirror segments, the segmentation trend is likely to increase in space. These complex apertures will have a strong impact on the coronagraph performance.

Different methods have been developed to correct for quasi-static aberrations with deformable mirrors (DMs). First, the correction for these errors along the full optical path requires a wavefront sensor operating directly in the science focal plane of the coronagraph. Several wavefront sensing techniques have been developed over the past decade, either using small hardware modifications \citep[e.g.][]{baudoz06, ndiaye16b} or by introducing known phase patterns in the pupil with the DMs \citep{borde06,giveon07,paul13,riggs16} to measure these errors. Secondly, using the output of these sensing techniques, high-contrast imaging requires specific correction algorithms to optimize the DM surfaces. For example, Stroke Minimization \citep[SM,][]{pueyo09} aims to reach a given contrast in a fixed region of an observed star image while minimizing the stroke applied to the DM actuators. This controlled high-contrast region in the focal plane of a coronagraph is referred to as dark hole (DH). Initially conceived for a single DM, these techniques have been improved to drive two sequential DMs, the first one in the pupil plane and the second one further along the beam propagation path \citep{pueyo07}. Indeed, while a single DM in a pupil plane can only modify the phase, an additional out-of-pupil-plane DM allows for the introduction of both phase and amplitude compensation and therefore, the generation of a symmetric DH around the target star image \citep{pueyo11,beaulieu17}. Most of these techniques have been tested experimentally in the presence of clear, circular apertures \citep{trauger07,mazoyer14}.

Simultaneously, several coronagraphs have already been designed to account for the effect of the central obscuration in the pupil, such as the Apodized Pupil Lyot Coronagraph \citep[APLC,][]{soummer03,soummer11,ndiaye15b}, the band-limited coronagraph \citep{kuchner02}, the Ring Apodized Vortex Coronagraph \citep[RAVC,][]{mawet05, mawet13} and the Phase-Induced Amplitude Apodization Complex Mask Coronagraph \citep[PIAACMC,][]{guyon2005, guyon10}. Finally, a vortex coronagraph with a new apodization design has been recently proposed, the Polynomial Apodized Vortex Coronagraph \citep[PAVC,][]{fogarty17} for apertures with central obscuration. Just like the RAVC, this apodization ensures perfect cancellation of an on-axis source in the Lyot-stop plane, but is less damaging in terms of off-axis source throughput. Finally, improved methods of apodization now allow the correction of non-axisymmetric discontinuities in the aperture (struts and segmentation of the primary). Among those, one can cite the last generation of APLCs \citep{ndiaye16}, of PIAACMCs \citep{Guyon14}, of shaped pupil coronagraphs \citep{carlotti13, zimmerman16} and of apodized vortex coronagraphs \citep{ruane15}.

In this paper, we introduce a new active technique to correct for aperture discontinuities, the Active Compensation of Aperture Discontinuities-Optimized Stroke Minimization (ACAD-OSM). In Section~\ref{sec:acad}, we present and compare the previous ACAD method \citep{pueyo13} and the new ACAD-OSM techniques. In Section~\ref{sec:compar_fix_apodiz}, we show that the performance of this method competes with one of state-of-the art static apodization coronagraphs designed for complex aperture \citep{ndiaye16}.

In Section~\ref{sec:aperture_disco}, we show that the ACAD-OSM can reach the $10^{-10}$ contrast a benchmark value corresponding to the contrast between Earth-like planets around Sun-like stars) with reasonable DM setups with any aperture and for 10\% bandwidth (BW). The impact of the amount and nature of the discontinuities on performance is also studied. In Section~\ref{sec:bandwidth}, the suitability of ACAD-OSM to produce high contrast DHs with larger BW (up to 30\%) is discussed. Finally, since this method is active, one of its main advantage is its ability to compensate for evolving or unknown optical aberrations or discontinuities in the pupil with the same algorithm. We finally show its capabilities to operate with optical design misalignments and/or phase errors in Sections~\ref{sec:misalignments} and~\ref{sec:phase_errors}.

%-------------------------------------------------------------------------------------------------
\begin{figure}
\begin{center}
 \includegraphics[width = .48\textwidth]{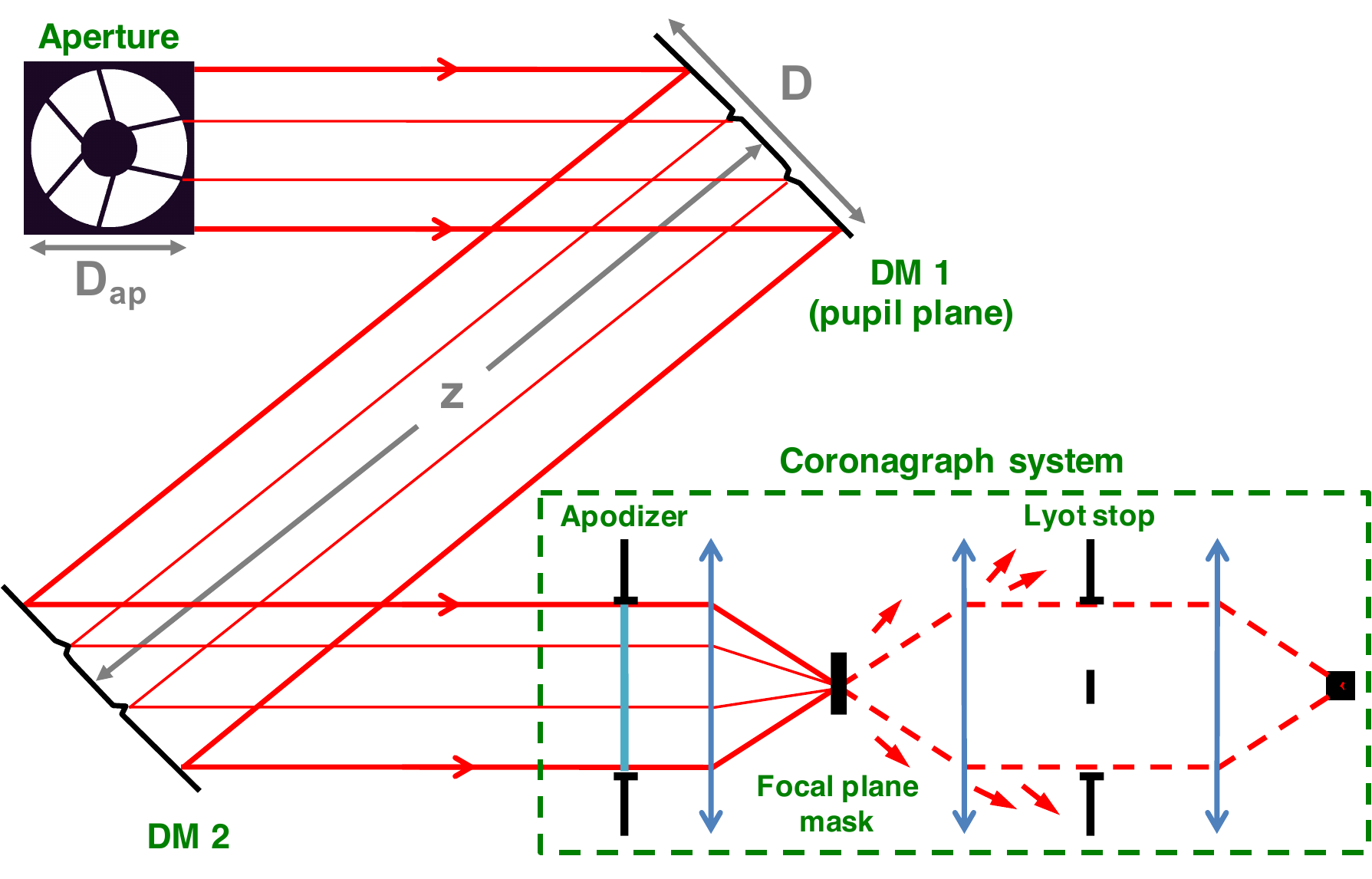}
 \end{center}
\caption[plop]
{\label{fig:schema_ACAD} Schematic representation of a two DM system and a coronagraph. The distances z, D and $D_{ap}$ are shown on this optical layout.}
\end{figure}
%-------------------------------------------------------------------------------------------------

Several parameters can be optimized for this technique to obtain the best performance on complex apertures. The exploration of the parameter space to optimize this method for future missions has been left for an upcoming paper \citep{mazoyer17b}, hereafter ACAD-OSM II.

%-------------------------------------------------------------------------------------------------
%-------------------------------------------------------------------------------------------------
%-------------------------------------------------------------------------------------------------
\section{The ACAD technique: initial goals and evolution}
\label{sec:acad}
%-------------------------------------------------------------------------------------------------
%-------------------------------------------------------------------------------------------------
%-------------------------------------------------------------------------------------------------

%-------------------------------------------------------------------------------------------------
\begin{table*}
\centering
\caption{This table shows the parameters of the two coronagraphs used in this article for different apertures.  Central obscuration and outer edge of the Lyot stop are defined as a percentage relative to the entrance pupil diameter $D_{ap}$. }
\label{tab:ularasa}
\begin{threeparttable}
\begin{tabular}{|c|c|c|c|c|c|}
\hline
              & FPM              & LS inner radius &  LS outer radius & Optimized for                & Transmission\\ \hline  \hline
\multicolumn{6}{|c|}{36\% central obscuration (WFIRST)}                                                                                                                                                               \\ \hline  \hline

PAVC6 & Charge 6 vortex    & 55\%         & 100\%              & ideal\tnote{$\dagger$}                                & 72\%                   \\ \hline  \hline
\multicolumn{6}{|c|}{17\% central obscuration (SCDA)}                                                                                                                                                                 \\ \hline  \hline
APLC          & Opaque Lyot mask: 4 $\lambda_0/D_{ap}$ radius & 30\%   & 92\%               & \begin{tabular}[c]{@{}c@{}}c = $10^{-10}$ (10\% BW)\end{tabular} & 58\%                   \\ \hline

PAVC6 & Charge 6 vortex    & 41\%        & 100\%            & ideal\tnote{$\dagger$}                                                          & 76\%                   \\ \hline \hline
\multicolumn{6}{|c|}{30\% central obscuration (E-ELT)}                                                                                                                                                                 \\ \hline  \hline
PAVC6 & Phase charge 6 vortex    & 39\%      & 100\%              & ideal\tnote{$\dagger$}                                                            & 78\%                   \\ \hline

\end{tabular}
\begin{tablenotes}
\item[$\dagger$] The coronagraph totally cancels the on-axis light at all wavelengths in the absence of wavefront aberrations or pupil discontinuities other than the central obscuration and for point-like stars.
\end{tablenotes}
\end{threeparttable}
\end{table*}
%-------------------------------------------------------------------------------------------------

%-------------------------------------------------------------------------------------------------
\subsection{Two DM aperture correction techniques}
\label{sec:acad_desc}
%-------------------------------------------------------------------------------------------------

Figure~\ref{fig:schema_ACAD} shows a schematic representation of the optical design used in this paper. The telescope aperture contains segment gaps from the primary mirror, as well as central obstruction and secondary mirror support structures. The aperture is circular in this paper, although this geometry is not a requirement for the technique described here. The beam is reflected on the first DM that is located in a pupil plane then propagates to the second DM, located at a distance $z$ further along in the beam path. After both reflections, a pupil plane is re-imaged, in which the coronagraph entrance pupil takes place with an apodizer. Downstream this part, there is a coronagraph optical layout, with a focal plane mask (FPM), a Lyot stop (LS) in the relayed pupil plane, and finally a detector in the final image plane.

The two DMs are assumed to be square with the same size $D\times D$, and the same number of actuators $N_{act}\times N_{act}$. $N_{act}$ is always 48 in this paper. The size $D$ is ofter write in the form $N_{act} *$ IAP, where IAP is the DM inter-actuator pitch. In this paper, the DM is $10\%$ larger than the circular aperture diameter $D_{ap}$. The region outside the second DM is also assumed to be reflective but not actuated, to mitigate the effects of potential clipping of the beam by the second DM. This effect is studied in details in ACAD-OSM II. Finally, $\lambda_0$ and $\Delta \lambda$ denote the central wavelength and width of the considered spectral BW. The central wavelength $\lambda_0$ is set at 550 nm in this paper. We assume null phase aberrations, except in Section~\ref{sec:phase_errors}. 

In this paper, this technique is only combined with two coronagraph designs that are optimized to correct for the effects of the pupil with central obscuration: the active correction is only used to address the other aperture discontinuities (segmentation and secondary struts). There is no fundamental difference between all the aperture features and one could also use mirror apodization to correct for the central obscuration effects. However, these effects involve large strokes that cannot be achieved with the state-of-the-art DM devices and are therefore out of the scope of this paper. Finally, this technique can also be associated with other coronagraph with more complex layout (e.g. PIAA).

The first coronagraph design used in this paper is the PAVC \citep{fogarty17} using a vortex FPM and an apodization. The vortex coronagraph is simulated using the numerical technique described in \citet[][]{mazoyer15}, in order to totally cancel the on-axis light in the Lyot stop plane at all wavelengths in the absence of wavefront aberrations or pupil discontinuities other than the central obscuration and for point-like stars. The second coronagraph design of this paper is the APLC \citep{soummer03,soummer11, ndiaye16}. The apodization, the opaque FPM radius, and the inner Lyot stop radius of this APLC have been optimized to maximize the throughput, while providing a $10^{-10}$ contrast over a 10$\%$ BW. Table~\ref{tab:ularasa} summarizes the parameters of the coronagraphs used in this paper.
 
%-------------------------------------------------------------------------------------------------
\subsection{The original ACAD solution}
\label{sec:acad_ros}
%-------------------------------------------------------------------------------------------------

The ACAD technique was introduced in \cite{pueyo13} with the goal to find the best DM shapes to correct for the effects of discontinuities in an aperture. In previous papers \citep{pueyo13, pueyo14, mazoyer15, mazoyer16b}, the ACAD method was presented in two successive steps: (i) an semi-analytic ray optic solution (hereafter ACAD-ROS) to flatten the electrical field in pupil plane, (ii) an SM algorithm \citep{pueyo09}, starting from the ACAD-ROS shapes and adding small adjustments to dig a DH in the focal plane.

\subsubsection{The ray optic solution}

ACAD-ROS is a semi-analytic solution to retrieve the two DM shapes to flatten the complex electrical field after a segmented aperture. This non linear technique is solving a Monge-Amp\`ere equation to find the optimal DM shapes. The applications of this technique far exceeds the field of the direct imaging of exoplanets. However, this technique has several important limitations for this specific application.

First, as showed in Section~\ref{sec:contrast_throughput}, this method gives limited results in contrast in challenging cases such as a large BW ($\Delta \lambda /\lambda_0 > 10$ \%) or large struts \citep{pueyo13, mazoyer16b}. Secondly, the DM strokes introduced in the ACAD-ROS step are proportional to the Fresnel number $D^2/\lambda z$ and range from a few hundreds of nanometers to a few micrometers. Using the HiCAT \citep[High-Contrast Imager for Complex Aperture Telescopes][]{ndiaye15} optical setup, we previously showed that the required ACAD-ROS solution strokes could barely be applied on Boston Micromachines (BMC\footnote{\url{http://www.bostonmicromachines.com/deformable-mirrors.html}}) DMs \citep{mazoyer16b}. More importantly, these very large DM actuator strokes from ACAD-ROS tend to scatter the light of the off-axis PSF at large separation, leading to an important effect on the throughput of a companion. Such a scattering of the PSF at large separation due to high strokes on the mirrors is one of the main limitations of static mirror apodization \citep{guyon2005}. Finally, because ACAD-ROS flatten semi-analytically the wavefront in the pupil plane, one practical disadvantage of this technique is that the solution cannot be applied to the DMs by only using a focal plane wavefront sensing technique.

%-------------------------------------------------------------------------------------------------
\begin{figure}
\begin{center}
 \includegraphics[width = .48\textwidth, trim=0.1cm 4.5cm 4.5cm 4cm, clip = true]{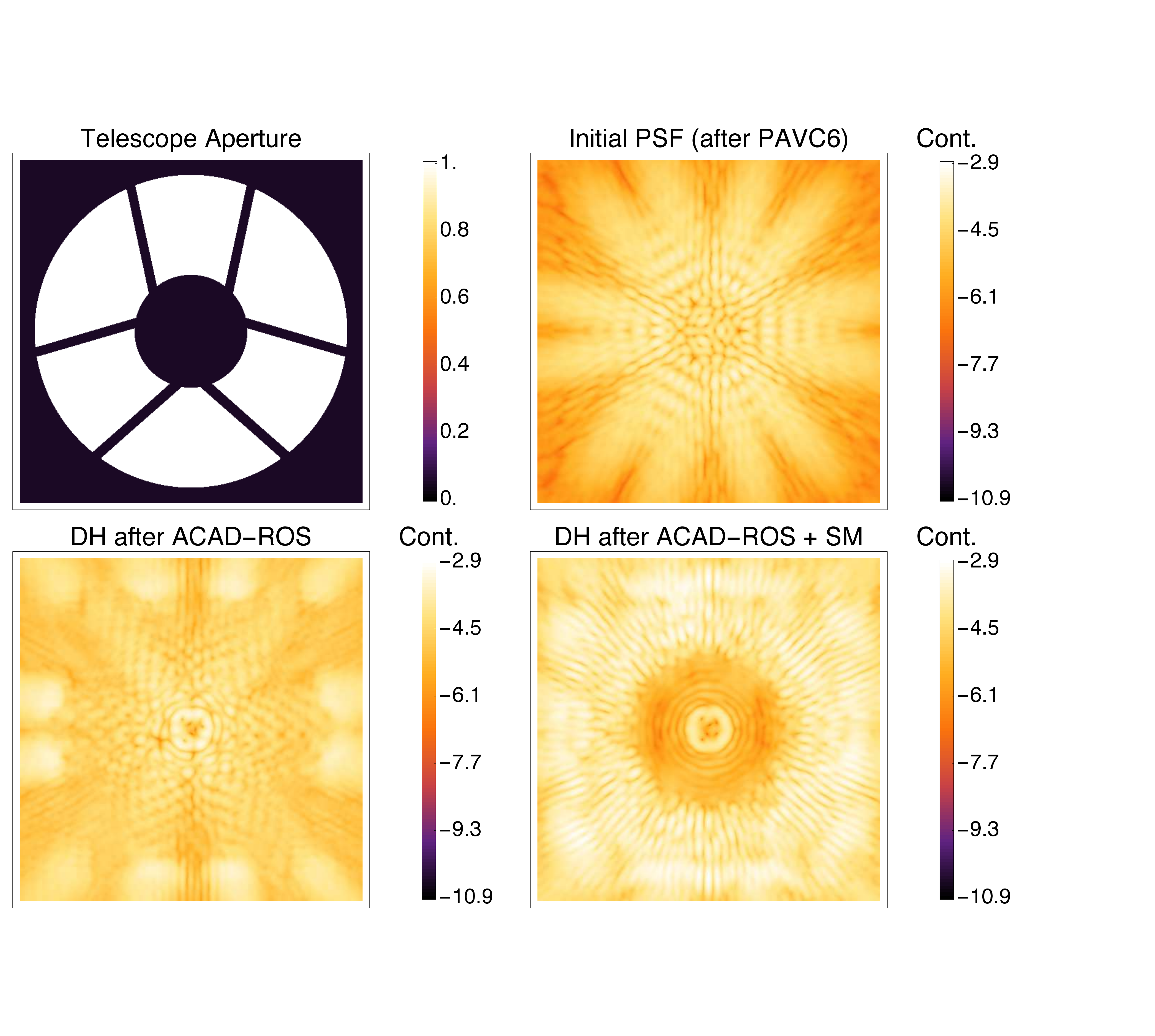}
 \end{center}
\caption[fig:wfirst_dh_ACADROS]
{\label{fig:wfirst_dh_ACADROS} Evolution of the PSF for the correction using ACAD-ROS followed by SM method (WFIRST aperture, charge 6 PAVC, $N_{act} = 48$, IAP = 1 mm, D = $48 * 1$ mm, $z = 1$ m, $\Delta \lambda /\lambda_0 = $ 10\% BW). Top, left: WFIRST aperture. Top right: the initial PSF after the coronagraph. Bottom, left: the DH created after the ACAD-ROS solution. Bottom, right: the final 3-10 $\lambda_0/D_{ap}$ DH created after ACAD-ROS and SM algorithms.}
\end{figure}
%-------------------------------------------------------------------------------------------------

%-------------------------------------------------------------------------------------------------
\begin{figure}
\begin{center}
 \includegraphics[width = .48\textwidth, trim= 0.1cm 4.5cm 4.5cm 4cm, clip = true]{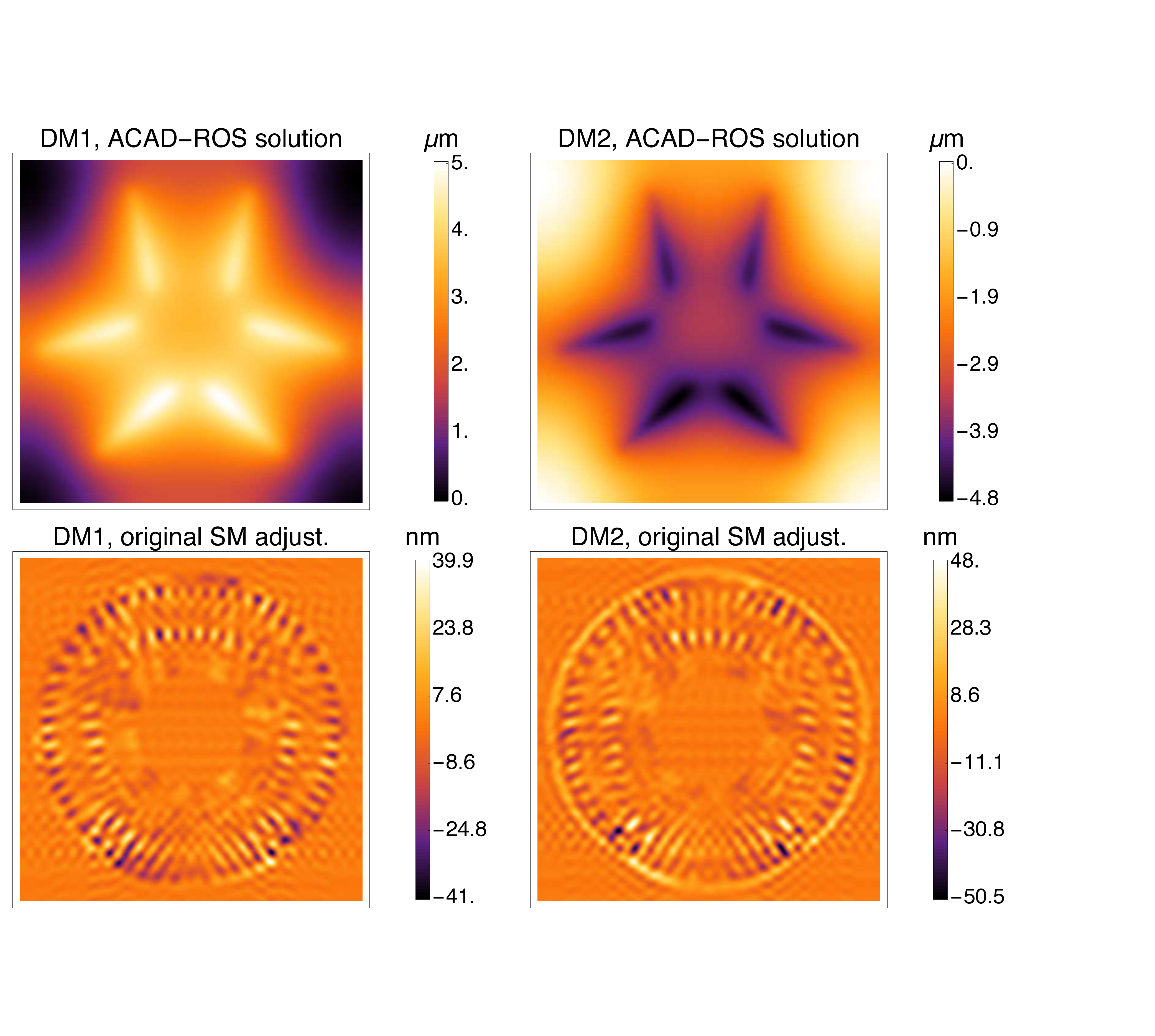}
 \end{center}
\caption[fig:wfirst_dm_ACADROS]
{\label{fig:wfirst_dm_ACADROS} DM shapes for the correction using ACAD-ROS followed by SM method (WFIRST aperture, charge 6 PAVC, $N_{act} = 48$, IAP = 1 mm, D = $48 * 1$ mm, $z = 1$ m, $\Delta \lambda /\lambda_0 = $ 10\% BW). Top: the two ACAD-ROS DM shapes. Bottom: DM shape adjustments after the SM step.}
\end{figure}
%-------------------------------------------------------------------------------------------------

%-------------------------------------------------------------------------------------------------
\subsubsection{The SM algorithm}
\label{sec:strokemin}
%-------------------------------------------------------------------------------------------------

Starting from the ACAD-ROS solution applied on the DMs, \cite{pueyo13} then propose to apply an SM algorithm to dig the DH. The SM algorithm \citep{pueyo09} is a correction technique for high-contrast imaging that can be used with single or several DM designs. It first uses a technique of estimation of the speckle complex electrical field in focal plane to build an interaction matrix. This interaction matrix links DM movements (described by a basis of specific movements) to their effect in the speckle complex electrical field estimation. The detailed construction of an interaction matrix can be found in \cite{mazoyer13c}. This matrix is then used to find the relevant DM shape(s) to correct for any speckle field due to phase and amplitude errors and produce a DH in the focal plane of a coronagraph. Like most correction algorithms, this method relies on an assumption of small aberrations relative to $\lambda_0$ to ensure that the propagation through the whole coronagraphic system is a linear function of the DM movements. This linear algorithm does not actually minimize the contrast but the strokes introduced on a DM to reach a target contrast $C_{target}$ in the DH. To ensure to find a local minimum in contrast with a small amount of strokes, the target contrast at each iteration $k$ is lowered, using the gain $\gamma$ of the algorithm :
\begin{equation}
\label{eq:gainSM}
C_{target}[k+1] = (1-\gamma) C_{target}[k]\,\,\,\,\, 0<\gamma<1\,\,.
\end{equation}
Larger gain goes with more aggressive algorithm, converging faster towards the minimum contrast but sometimes leading to oscillations, or even divergence. 

\subsubsection{Results with the WFIRST aperture}

The results of the combination of ACAD-ROS and SM for the WFIRST aperture are shown in Figs.~\ref{fig:wfirst_dh_ACADROS} and \ref{fig:wfirst_dm_ACADROS}. The coronagraph is a charge 6 PAVC, with the DM setup chosen for this mission ($N_{act} = 48$, IAP = 1 mm, D = $48 * 1$ mm, $z = 1$ m). Finally, the BW is 10\%. First, the two DM shapes are analytically retrieved using ACAD-ROS method (Fig.~\ref{fig:wfirst_dm_ACADROS} top panels). These shapes, with $\sim 5 \mu$m strokes (peak-to-valley), are remapping the field to average the effects of the struts inside the DH (Fig.~\ref{fig:wfirst_dh_ACADROS} bottom left panel). Finally, using these shapes as the initial states, the SM algorithm is applied to produce DM adjustments of a few tens of nanometers (Fig.~\ref{fig:wfirst_dm_ACADROS} bottom panels) on top of the ACAD-ROS shapes. The 3-10 $\lambda_0/D_{ap}$ DH generated in the focal plane is shown in Fig.~\ref{fig:wfirst_dh_ACADROS} (bottom right panel).

%-------------------------------------------------------------------------------------------------
\subsection{The new adaptive interaction matrix method: ACAD-OSM algorithm}
\label{sec:acad_osm_desc}
%-------------------------------------------------------------------------------------------------

%-------------------------------------------------------------------------------------------------
\begin{figure}
\begin{center}
 \includegraphics[width = .3\textwidth, clip = true]{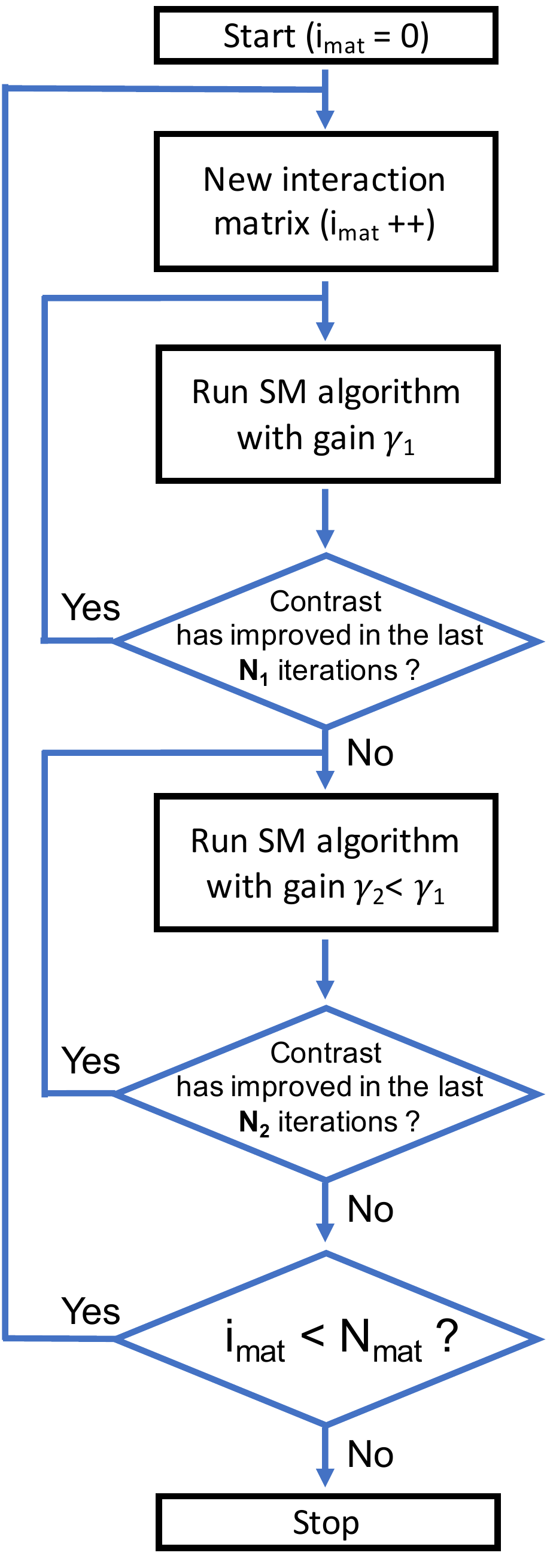}
 \end{center}
\caption[fig:block_diagram_acad]
{\label{fig:block_diagram_acad} Block diagram of the ACAD-OSM algorithm.}
\end{figure}
%-------------------------------------------------------------------------------------------------

This paper introduces a new active technique for the correction of aperture discontinuities: ACAD-OSM. As in \cite{pueyo13}, a perfect way to estimate the complex speckle electrical field in the focal plane at each wavelength is assumed, and this paper focus on the correction algorithm only. The SM algorithm shows good results for the correction of small phase and amplitude errors but it diverges quickly for the correction of aperture discontinuities. This divergence occurs when the DM strokes approach $\lambda/5$. This behavior was expected since these strokes are too large to assume a linear relation between the DM movements and their effects in the focal plane. Indeed, aperture discontinuities correction usually require far larger strokes than ``small'' phase and amplitude errors.

The optimized stroke minimization algorithm is built to avoid this problem, shown in the form of a block diagram in Fig~\ref{fig:block_diagram_acad}. First, it uses a variable SM loop gain $\gamma$ define in Eq.~\ref{eq:gainSM} to set the target image intensity at each iteration of the SM algorithm. For the first iterations in the linear range of the initial DM shapes, the gain is set high ($\gamma = \gamma_1$) to converge faster toward the best solution. When the algorithm starts to diverge, the DM shapes derived from the linear SM algorithm do not improve the DH contrast anymore. Consequently, after $N_1$ iterations without improvement, the SM gain is decreased in an attempt to push the correction forward: $\gamma = \gamma_2$. Secondly, if the contrast still does not improve with the new gain after a given number of iterations $N_2$, the limit of the linearity range allowed by the initial DM shapes has been reached. In this case, the linearity range is re-centered around the final DM shapes from the previous step that are set as the new reference level to build a new interaction matrix. This operation is repeated $N_{mat}$ times. In this paper, $N_1 =10$, $N_2 =20$, $\gamma_1 = 5\%$ and $\gamma_2 = 2\%$. 

With every matrix re-computation, the contrast improvement tends to decrease, implying that the limitation of the convergence is not anymore the linearity range but the proximity of a local minimum. We usually stop the algorithm after the eighth matrix ($N_{mat} = 8$), since in most cases the additional contrast improvement is less than a factor of 2 at that point. In some rare cases (requiring large strokes on the DMs), more matrices are generated (usually $N_{mat} = 10$, sometimes $N_{mat} = 12$) to make sure that the local minimum is reached. For the simplest correction situations (monochromatic, more friendly apertures), the number of matrices before reaching a local minimum can usually be set smaller ($N_{mat}\sim 5$).

At a first glance, this algorithm may seem complicated to apply in real life but one can notice that most high-contrast testbeds are already using a similar strategy to achieve high contrast. Indeed, correction usually starts from a random state of the DM surface in which the phase errors can be as large as a few tens of nm. Experimentally, a few interaction matrices are usually necessary to reach the sub-nanometric error level. The algorithm presented here follows the same experimental approach, except that the strokes to correct for the aperture discontinuities are more important and therefore require more iteration steps.

\begin{figure}
\begin{center}
 \includegraphics[width = .48\textwidth, trim= 0.1cm 4.5cm 4.5cm 4cm, clip = true]{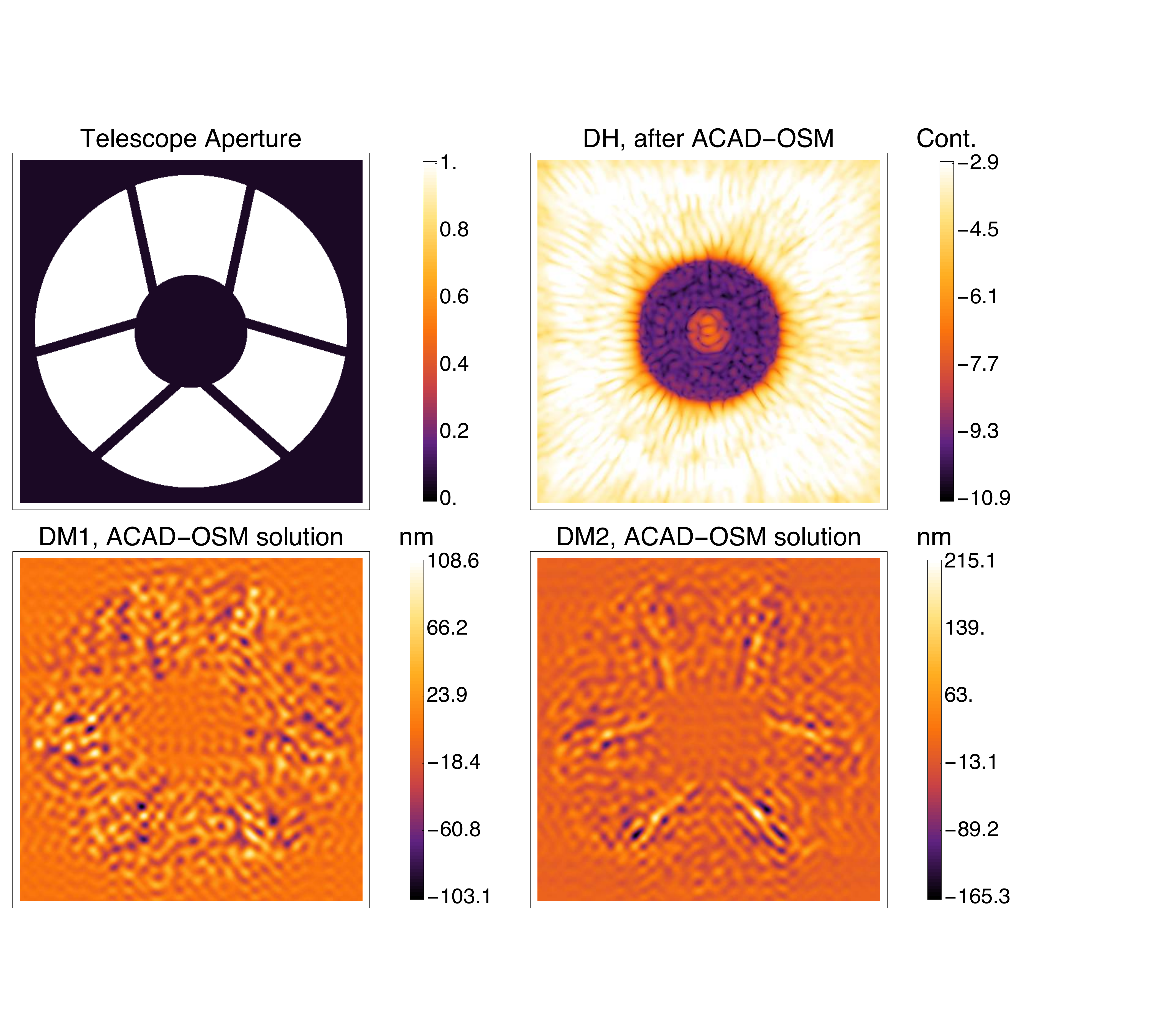}
 \end{center}
\caption[fig:wfirst_dh_ACADOSM]
{\label{fig:wfirst_dh_ACADOSM} WFIRST aperture with the ACAD-OSM technique (charge 6 PAVC, $N_{act} = 48$, IAP = 1 mm, D = $48 * 1$ mm, $z = 1$ m, $\Delta \lambda /\lambda_0 = $ 10\% BW). Top left: WFIRST aperture. Top right: the final 3-10 $\lambda_0/D_{ap}$ DH. Bottom: the DM shapes obtained using ACAD-OSM.}
\end{figure}

Fig~\ref{fig:wfirst_dh_ACADOSM} shows the result of ACAD-OSM algorithm with the WFIRST aperture with 8 consecutive matrices. The 2 DMS shapes are shown on the bottom and the final DH is shown on the top right. We compare these results with the results obtained with ACAD-ROS + SM in the next section.

%-------------------------------------------------------------------------------------------------
%-------------------------------------------------------------------------------------------------

%-------------------------------------------------------------------------------------------------
\subsection{Comparison of the two techniques}
\label{sec:compar_ROS_OSM}
%-------------------------------------------------------------------------------------------------

%-------------------------------------------------------------------------------------------------
\subsubsection{Description of the performance metrics}
\label{sec:contrast_throughput}
%-------------------------------------------------------------------------------------------------

All the results in this paper have been obtained following the same method. In the correction process of ACAD-OSM, the interaction matrix is built by concatenating interaction matrices at several wavelengths (3 or more). However, once the DM shape solutions are obtained, a larger number of simulated wavelengths within the operating bandpass is used to produce the final broadband image on which the results are measured (usually one sampling bandwidth per percent of the bandwidth). The following metrics are used in this paper:

\begin{itemize}
    \item \textit{Contrast}. In this paper, the contrast is defined as the amount of star light at a given point of the focal plane normalized to the intensity peak of the on-axis PSF in the final focal plane and in the absence of the coronagraph FPM. Contrast curves are usually plotted in this paper, i.e. azimuthally averaged intensity profiles in the focal plane as a function of the distance to the star image (in telescope resolution elements $\lambda_0/D_{ap}$). Finally, the performance of a solution is sometimes presented using only the averaged DH contrast. This last indicator is imperfect but useful to quickly compare results. 
    
    \item \textit{Throughput}. To capture the complex effects of the ACAD-OSM technique on an off-axis PSF, the throughput is defined: for an off-axis source at a given separation from the star, this value represents the ratio of the energy reaches the off-axis PSF core in the final image plane (inside a photometric aperture of 0.7 $\lambda_0/D_{ap}$ radius centered around the expected position of the PSF) to the energy in the on-axis PSF core (inside the same photometric aperture) when the 2 DM and the coronagraph systems are removed. The throughput takes into account both the apodization transmission and the scattering of the off-axis PSF due to the propagation through non-clear pupils and non-flat DMs. Only throughput losses due to the two DM system and the coronagraph throughput are taken into account: all other factors of throughput (including all mirror reflectivities, detector quantum efficiency, polarizers) are neglected. In the following, performance are usually shown using curves showing the throughput value as a function of the off-axis source separation in $\lambda_0/D_{ap}$.
    
\end{itemize}

%-------------------------------------------------------------------------------------------------
\begin{figure*}
\begin{center}
 \includegraphics[width = .49\textwidth, trim= 1.2cm 0.8cm 0.7cm 0.3cm, clip = true]{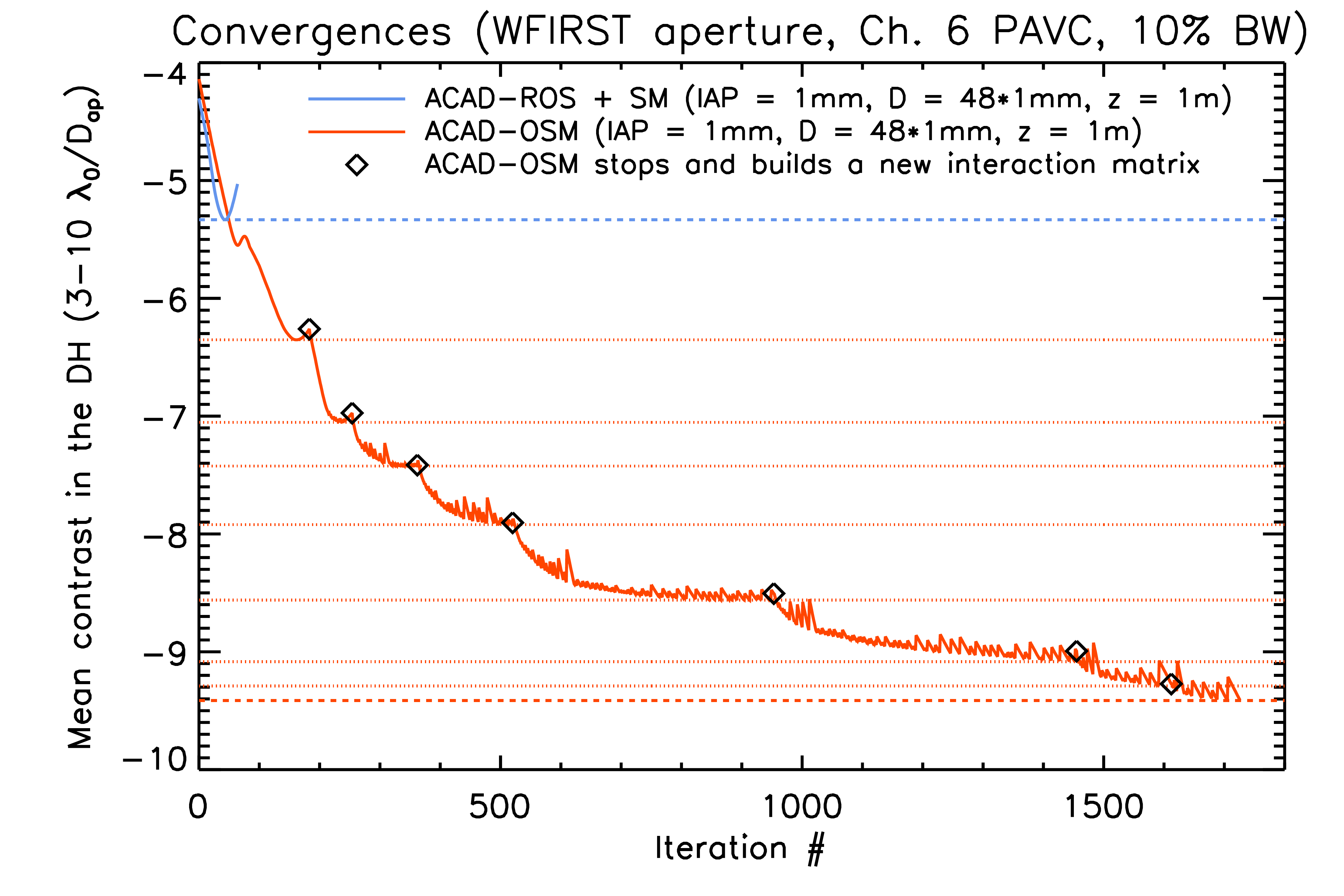}
 \includegraphics[width = .49\textwidth, trim= 1.2cm 0.8cm 0.7cm 0.3cm, clip = true]{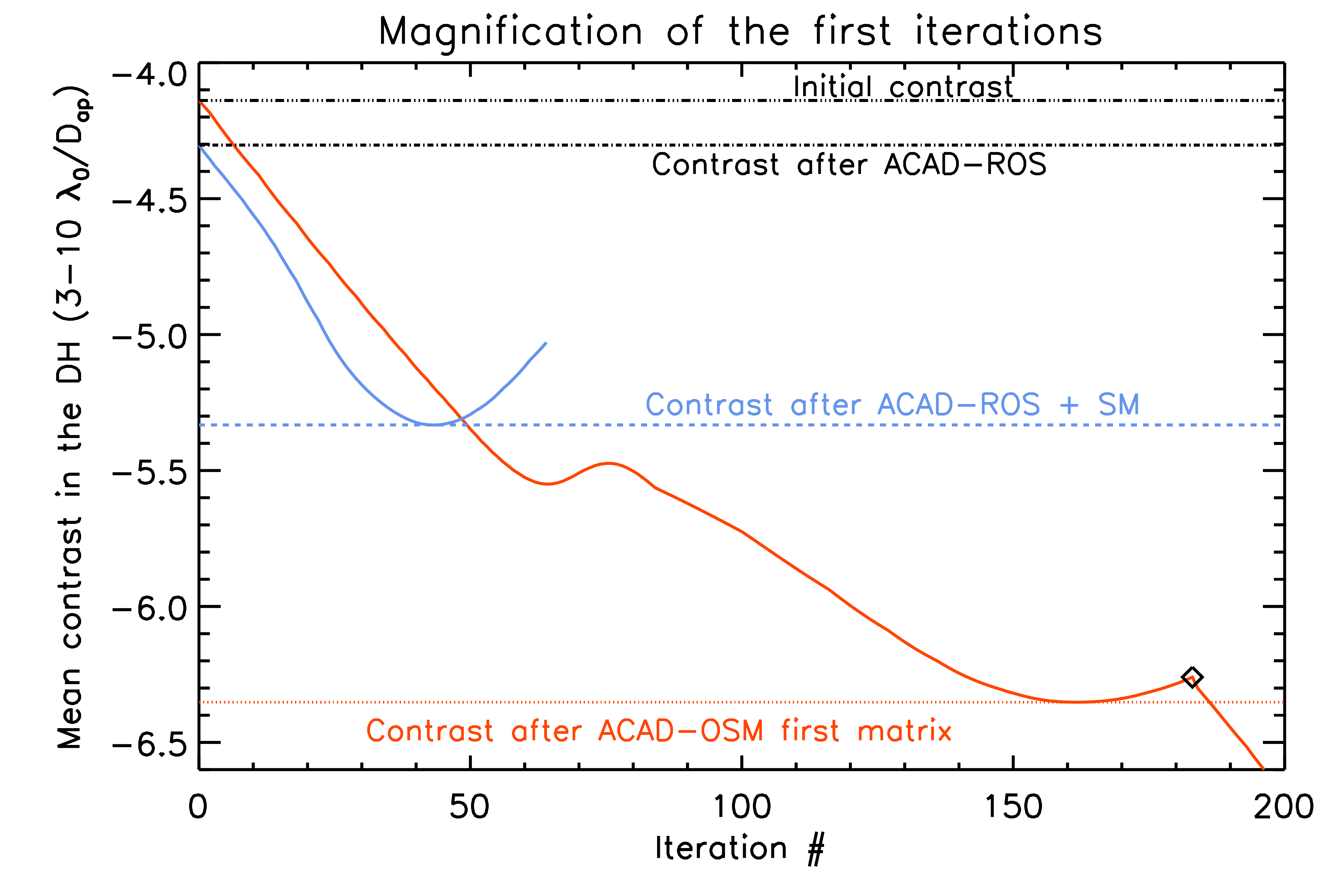}
 \end{center}
\caption[fig:Iter_wfirst]
{\label{fig:Iter_wfirst} Convergence of the mean contrast level in the DH as a function of the number of iterations for the ACAD-ROS + SM solution (blue curve) and for the ACAD-OSM solution in 8 matrices (red curve). The coronagraph is a charge 6 PAVC and the DM setup is the one used in the WFIRST mission: WFIRST aperture, $N_{act} = 48$, IAP = 1 mm, D = $48*1$ mm, $z = 1$ m, $\Delta \lambda /\lambda_0 = $ 10\% BW. The black diamonds are located on the iterations where the algorithm stops and recalibrates with a new interaction matrix. At each step, the contrast level is indicated by a horizontal dotted line. Final contrasts for both methods are shown with dash lines. The right plot shows a magnification of the first 200 iterations from the left plot.}
\end{figure*}
%-------------------------------------------------------------------------------------------------

%-------------------------------------------------------------------------------------------------
\subsubsection{Comparison}
\label{sec:compar_ROS_OSM_results}
%-------------------------------------------------------------------------------------------------

Fig.~\ref{fig:Iter_wfirst} (left) shows the mean contrast level in the 3-10 $\lambda_0/D_{ap}$ DH as a function of the iteration number with ACAD-ROS+SM (in blue) and ACAD-OSM (in red) methods. For ACAD-OSM, the 8 black diamonds show the iterations at which the algorithm stops and builds a new interaction matrix. The red horizontal dotted lines represent the best contrast achieved with each interaction matrix. Fig.~\ref{fig:Iter_wfirst} (right) plot shows a magnification of the first 200 iterations. The black horizontal lines shows the initial contrasts for both methods (PSF shown in Fig.~\ref{fig:wfirst_dh_ACADROS} top right panel) and the contrast level after the ACAD-ROS (PSF shown in Fig.~\ref{fig:wfirst_dh_ACADROS} bottom left). The thicker dashed lines represent the final contrast in both cases: blue after ACAD-ROS and SM (DH in Fig.~\ref{fig:wfirst_dh_ACADROS} bottom right) and red for ACAD-OSM (DH in Fig.~\ref{fig:wfirst_dh_ACADOSM} top right). 

The initial DH contrast level with initial flat DM shapes and after ACAD-ROS is $10^{-4.14}$ and $10^{-4.3}$, showing an improvement of only 1.4 (however, with easier apertures or DM setups \cite{pueyo13} reported better improvements). In the first 50 iterations, the ACAD-ROS and SM method shows better contrast than the ACAD-OSM. However, the SM algorithm after ACAD-ROS diverges after only 40 iterations, due to the fact that the SM enters into a non-linear regime. The final contrast with ACAD-ROS+SM method is $10^{-5.33}$ (horizontal blue dashed line). At about 60 iterations, the ACAD-OSM algorithm also diverges. At this moment, it switches to the low-gain mode, allowing it to continue converging for 110 more iterations after which it diverges again at about 185 iterations. At this point, the limit of linearity for this matrix is reached. With an intermediary contrast of $10^{-6.35}$, a new matrix is built and the correction continue to progress towards the local contrast minimum. After running the algorithm for eight interaction matrices, the final mean contrast (Fig.~\ref{fig:wfirst_dh_ACADOSM} top right) is now $10^{-9.41}$ (horizontal red dashed line). The improvement in contrast between the seventh and the eighth matrices is only a factor 1.4. 

\begin{figure}
\begin{center}
 \includegraphics[width = .48\textwidth,trim= 1.1cm 0.8cm 0.7cm 0.3cm, clip = true]{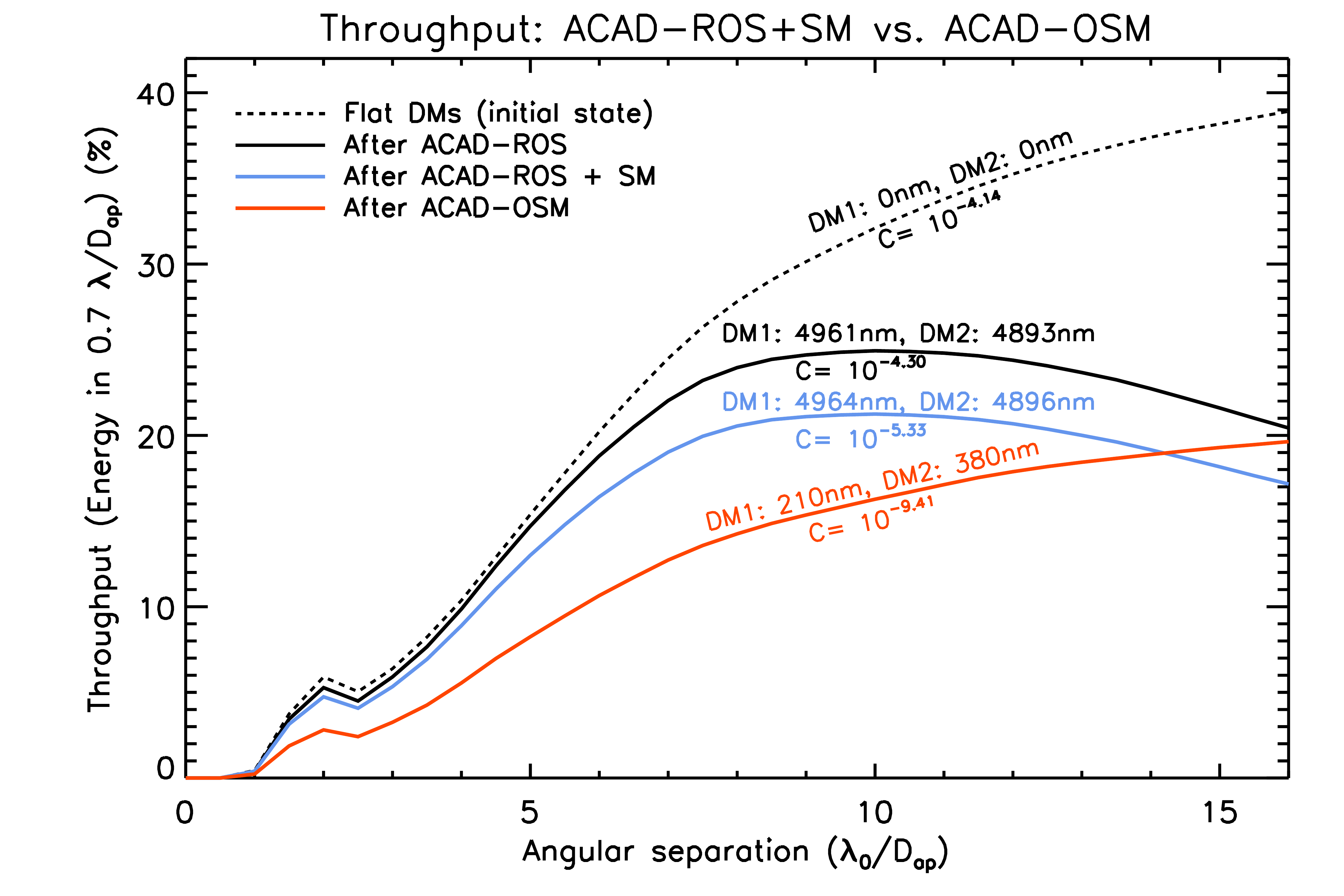}
 \end{center}
\caption[plop]
{\label{fig:throughput_compar_inner_acad_acadosm} Throughput results for the WFIRST aperture with flat DMs (black dashed line), then after the ACAD-ROS shapes are applied on the DMs (black solid line), after ACAD-ROS and SM (blue curve) and finally for the ACAD-OSM solution in 8 matrices (red curve). The coronagraph is a charge 6 PAVC and the DM setup is the one used in the WFIRST mission: WFIRST aperture, $N_{act} = 48$, IAP = 1 mm, D = $48*1$ mm, $z = 1$ m, $\Delta \lambda /\lambda_0 = $ 10\% BW.}
\end{figure}

The performance in throughput are shown in Fig.~\ref{fig:throughput_compar_inner_acad_acadosm}. For each solution, we note the associated performance in contrast and the strokes (peak-to-valley) on the DMs. The ACAD-ROS algorithm have surprisingly very little impact on the throughput at small separations (with strokes of $\sim 5~\mu$m), but starts to degrades drastically the throughput after 10 $\lambda_0/D_{ap}$ only. This throughput drop can be observed in all ACAD-ROS solutions \citep[e.g.][]{mazoyer15}. In comparison, the effect of SM and ACAD-OSM algorithms on throughput is more uniform over the focal plane. Finally, the throughput is better for the ACAD-ROS + SM algorithm than for the ACAD-OSM for the 0-14 $\lambda_0/D_{ap}$ range but quickly drops after. 

Here, the starting point of the ACAD-OSM method is the flat DM shapes. We also considered applying ACAD-OSM starting from the ACAD-ROS shapes, leading to close results in contrast in both situations. However, the throughput results always shows a quick drop at large separation with the ACAD-ROS technique, that does not appear with flat initial DMs. The number of matrices used in the ACAD-OSM have an influence on both throughput and contrast, that will be studied in ACAD-OSM II.

%-------------------------------------------------------------------------------------------------
%-------------------------------------------------------------------------------------------------

This section showed that ACAD-OSM shows better results both in contrast and in throughput at large separation than the previously developed active method. In the next section, its performance are compared to the one of state-of-the-art static apodization coronagraph designs optimized for complex aperture.

%-------------------------------------------------------------------------------------------------
\section{Comparison with static apodization coronagraphs}
\label{sec:compar_fix_apodiz}
%-------------------------------------------------------------------------------------------------

Several coronagraphs have been designed to obtain high performance with any kind of aperture, with an optimized apodizer. In this section, we compare the results of an APLC with an apodizer optimized for the full aperture (central obstruction, secondary structures and segmentation) \citep{ndiaye16}, and an APLC with an apodizer only optimized for the centrally obstructed pupil, used in combination with a two DM setup using the ACAD-OSM algorithm to correct for the secondary structures and the segmentation. For this test, a more friendly aperture is used, currently under study for future space telescopes in the Segmented Coronagraph Design and Analysis (SCDA) program\footnote{This research program is led by NASA's Exoplanet Exploration Program (ExEP). Please check: \url{https://exoplanets.nasa.gov/system/internal_resources/details/original/211_SCDAApertureDocument050416.pdf}}. This pupil, shown on Figure~\ref{fig:scda_dh} (top, left), hereafter SCDA aperture, includes a central obscuration (17\% of the aperture outer radius), finer secondary structures (0.6\%  of the left area), and a primary mirror with hexagonal segments.

%-------------------------------------------------------------------------------------------------
\begin{figure}
\begin{center}
 \includegraphics[trim= 1.5cm 1cm 1.0cm 0.5cm, clip = true,width = .48\textwidth]{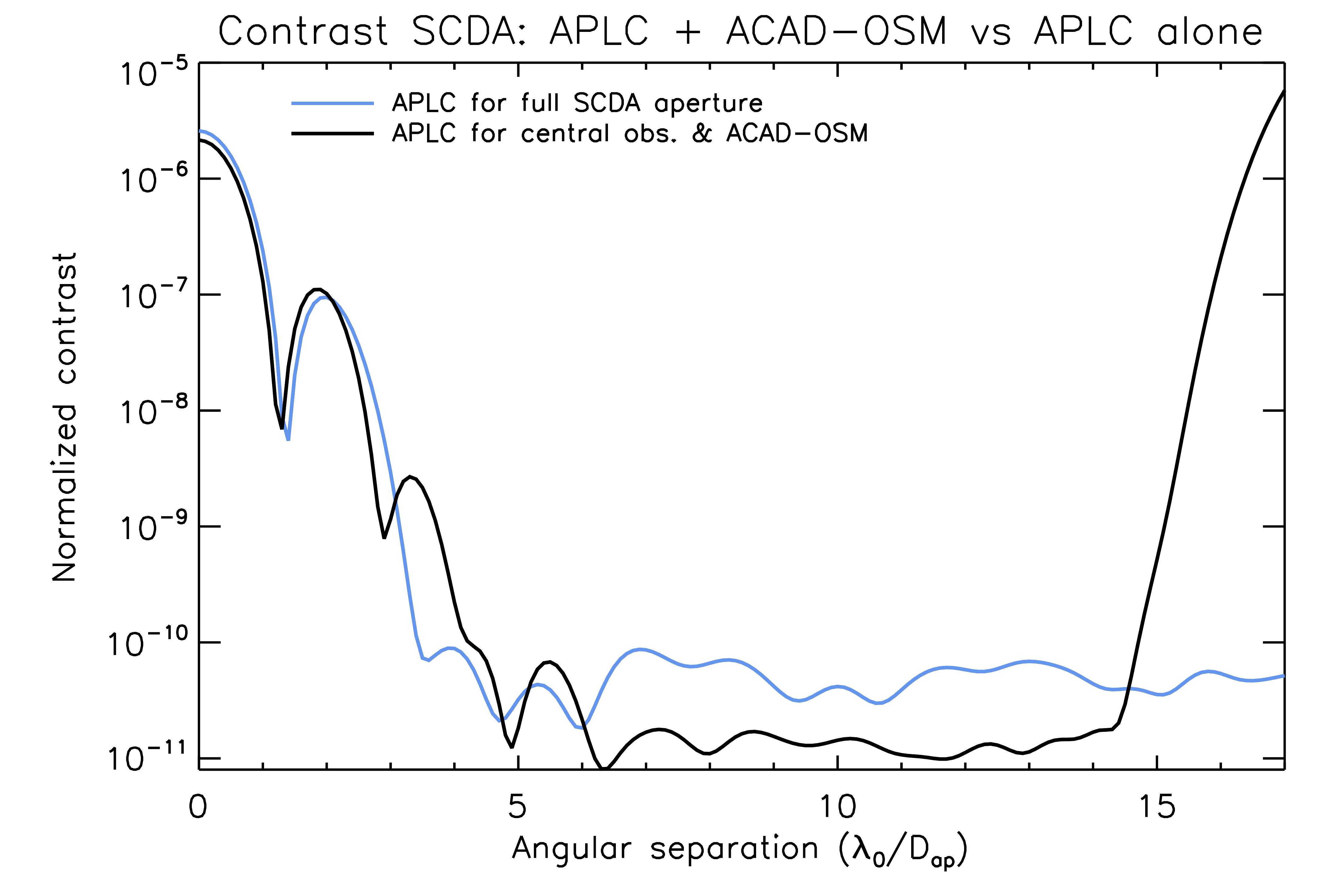}
  \end{center}
\caption[fig:comparAPLC_contrast]
{\label{fig:comparAPLC_contrast} Contrast results for the SCDA aperture and different versions of the APLC. In blue are show the performance for an APLC with an apodizer optimized for the full aperture. In black is shown the results for an APLC with an apodizer that is only optimized for the central obscuration combined with an ACAD-OSM system. The DM setup is $N_{act} = 48$, IAP = 0.3 mm, D = $48 * 0.3$ mm, $z = 0.3$ m, and the BW is $\Delta \lambda /\lambda_0 = $ 10\%.}
\end{figure}
%-------------------------------------------------------------------------------------------------

The performance in contrast (Fig.~\ref{fig:comparAPLC_contrast}) are similar, with better contrast in the 6-14 for APLC+ACAD-OSM and better contrast in the 3-5 $\lambda/D_{ap}$ range for full aperture-optimized APLC. However, only a small DH is corrected compared to the one created by the aperture-optimized APLC. Indeed, with DMs with only 48 actuators in the principal directions, the achievable frequency is limited to less than 24 $\lambda_0/D_{ap}$.

%-------------------------------------------------------------------------------------------------
\begin{figure}
\begin{center}
 \includegraphics[trim= 2.5cm 1cm 1.0cm 0.5cm, clip = true,width = .48\textwidth]{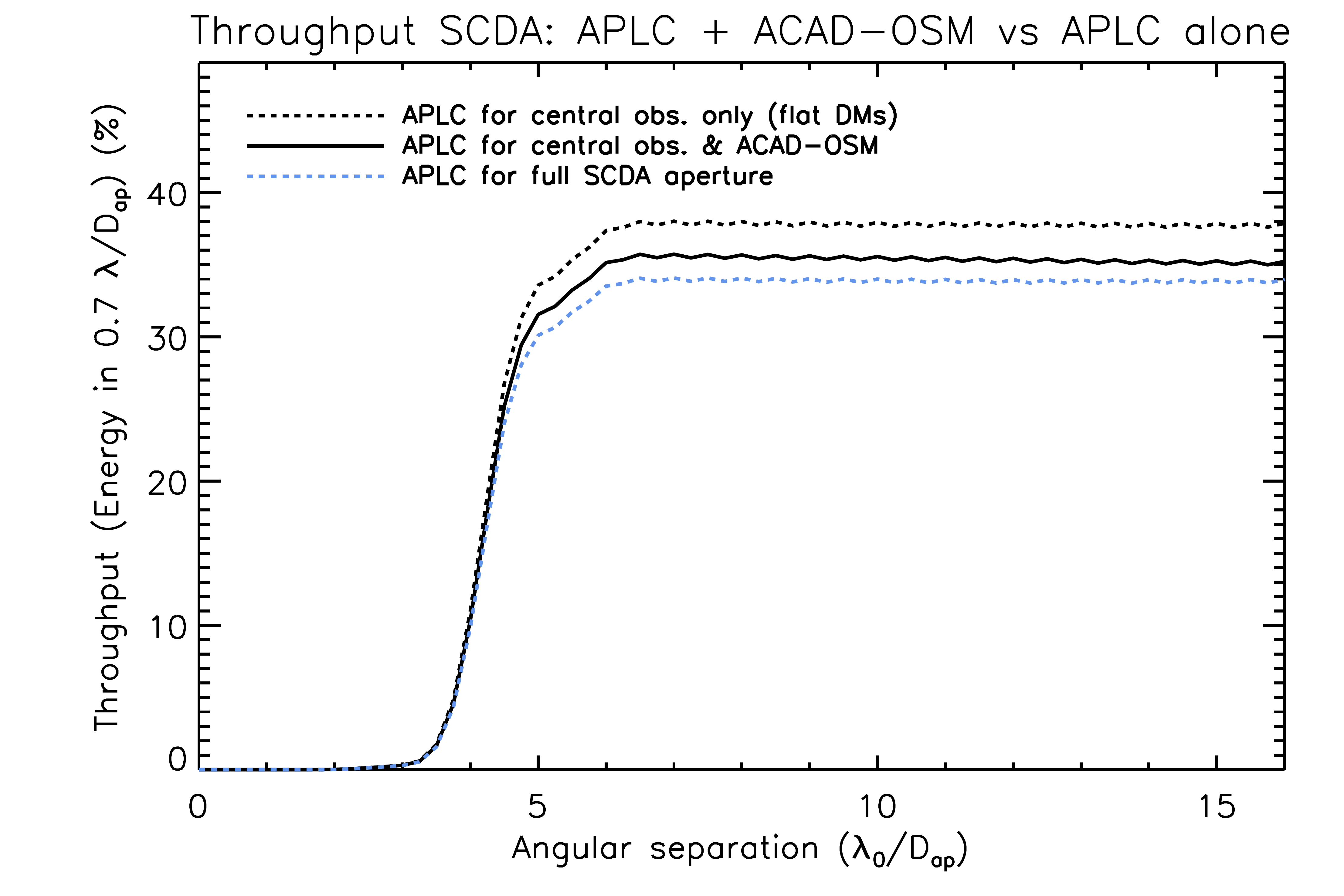}
  \end{center}
\caption[fig:comparAPLC_throughput]
{\label{fig:comparAPLC_throughput} Throughput results for the SCDA aperture and different versions of the APLC. The dashed lines show the throughput with flat DMs, before any correction (due to the APLC alone). In blue are show the results in contrast for a full aperture-optimized APLC (central obstruction, secondary structures and segmentation). In black is shown the results for an APLC with an apodizer that is only optimized for the central obscuration in addition to a ACAD-OSM system. The DM setup is $N_{act} = 48$, IAP = 0.3 mm, D = $48 * 0.3$ mm, $z = 0.3$ m, and the BW is $\Delta \lambda /\lambda_0 = $ 10\%.}
\end{figure}
%-------------------------------------------------------------------------------------------------

Fig.~\ref{fig:comparAPLC_throughput} shows the results in throughput. The two dashed curves show the results of the two apodizers, before any correction, with flat DMs: APLC for the central obscuration alone (black dashed line) vs full aperture-optimized APLC (central obstruction, secondary structures and segmentation) in a blue dashed line. The apodizer for the central apodization only is obviously better for the throughput. The result of a central apodization-optimized APLC, combined with an ACAD-OSM system is shown with a black solid line. The ACAD-OSM solution has a slightly higher throughput than the static, full aperture-optimized APLC. ACAD-OSM II shows that this DM setup is favorable to the correction reaching high contrast and low throughput: a less optimal DM setup could actually degrade performance.

In this section, we showed that the performance in term of throughput and contrast of the ACAD-OSM technique can compete with state-of-the-art static full aperture apodization techniques.

%-------------------------------------------------------------------------------------------------
%-------------------------------------------------------------------------------------------------
%-------------------------------------------------------------------------------------------------
\section{Impact of aperture discontinuities on the results of the ACAD-OSM method}
\label{sec:aperture_disco}
%-------------------------------------------------------------------------------------------------
%-------------------------------------------------------------------------------------------------
%-------------------------------------------------------------------------------------------------
In this section, we study the impact of several aperture discontinuity features on ACAD-OSM performance. The same coronagraph (charge 6 PAVC), DM setup ($N_{act} = 48$, IAP = 0.3 mm, D = $48 * 0.3$ mm, $z = 0.3$ m) and BW ($\Delta \lambda /\lambda_0 = $ 10\%) are used in this section.

%-------------------------------------------------------------------------------------------------
\begin{figure}
\begin{center}
 \includegraphics[trim= 1.5cm 0cm 0cm 0cm, clip = true,width = .48\textwidth]{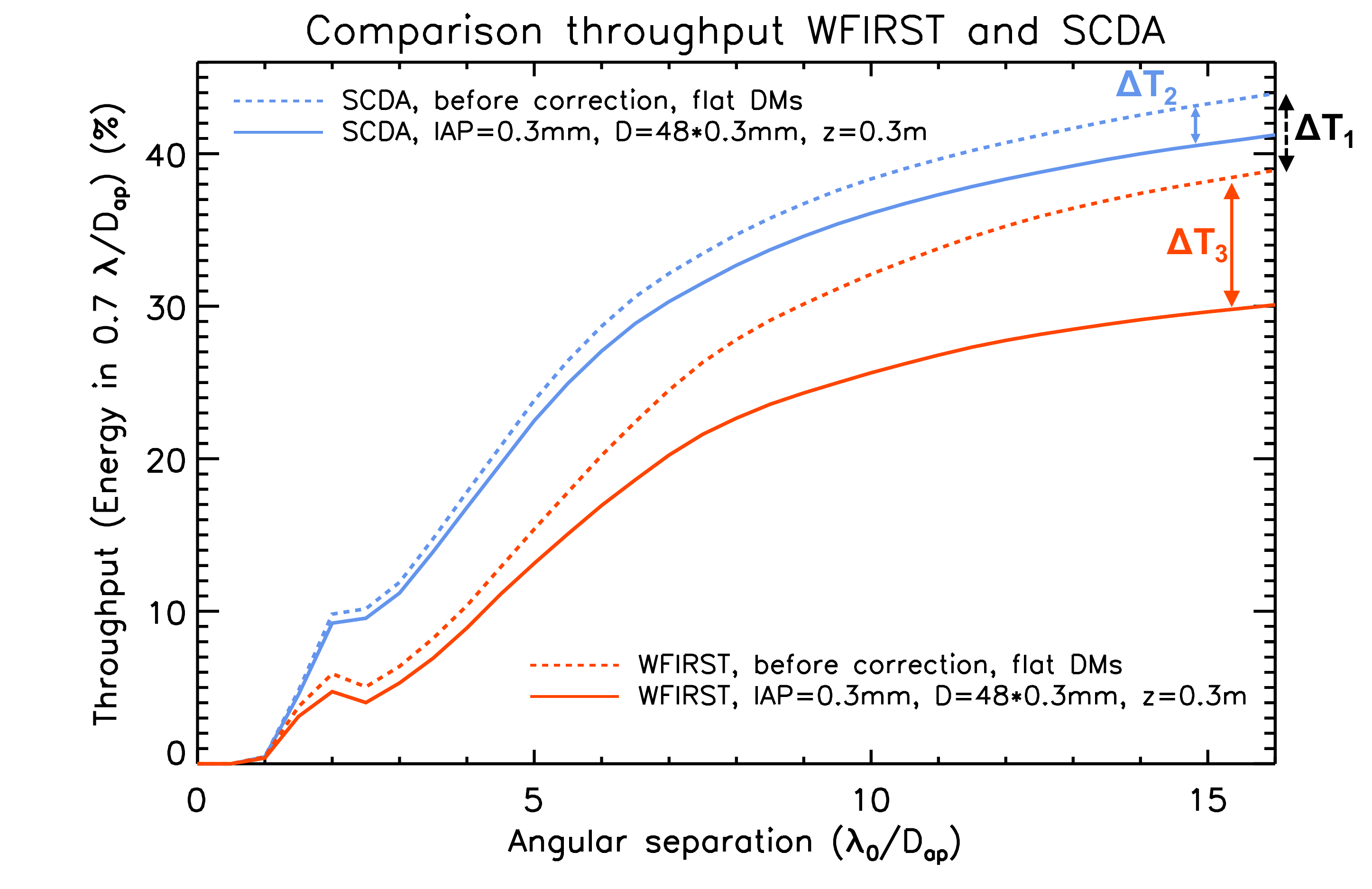}
  \end{center}
\caption[fig:wfirst_scda_throughput_compar_arrows]
{\label{fig:wfirst_scda_throughput_compar_arrows} Throughput results for different apertures, for the same DM setup and a charge 6 PAVC. The dashed lines show the throughput with flat DMs, before any correction (due to the PAVC alone). The solid lines show the throughput after ACAD-OSM corrections ($N_{act} = 48$, IAP = 0.3 mm, D = $48 * 0.3$ mm, $z = 0.3$ m and $\Delta \lambda /\lambda_0 = $ 10\%). The blue lines show the results for the SCDA aperture, and the red lines for the WFIRST aperture.}
\end{figure}
%-------------------------------------------------------------------------------------------------

\subsection{Central obscuration}

In this paper, all apodizers have been optimized to reach a $10^{-10}$ contrast (APLC) or better (PAVC) with a central obscuration therefore we do not study its influence on the contrast.
Fig.~\ref{fig:wfirst_scda_throughput_compar_arrows} shows the impact of the pupil discontinuities on off-axis throughput. The throughput before any correction (with flat DMS, dashed lines) and after the ACAD-OSM technique (solid lines) is shown for the SCDA aperture (blue lines) and the WFIRST aperture (red lines). 

The difference between the results in throughput for the two apertures before any correction is applied to the DM (difference between the dashed curves) is equal to $\Delta T_1$ = 5\% at 16 $\lambda_0/D_{ap}$. This difference is not due to the technique described here and is only related to the central obscuration size and to the method of the coronagraph apodization (PAVC here). Indeed, an increase in the central obscuration usually induces a decrease in the off-axis throughput \citep{ndiaye15b,fogarty17}. 

However, non-axisymmetric aperture discontinuities have a important impact on the performance of the ACAD-OSM technique, and are studied in in the next sections.

\subsection{Segmentation and struts}

The influence of aperture discontinuities on the contrast level is mainly driven by the widths of the struts. For a given number of actuators, the SCDA aperture gives better contrast than the WFIRST aperture. For example, the ACAD-OSM technique obtains a result of $10^{-11.2}$ between 1.5 and 15 $\lambda_0/D_{ap}$ with the SCDA aperture (contrast curve shown in magenta in Fig.~\ref{fig:bw_contrast}) and a contrast of $10^{-10.6}$ between 3 and 10 $\lambda_0/D_{ap}$ for the WFIRST aperture (contrast curve not shown here). 

For the same DM setup, the correction of large struts have a more important influence on the throughput, due to the larger strokes required on the DMs that can have a severe impact on the shape of the PSF at large separations. In Fig.~\ref{fig:wfirst_scda_throughput_compar_arrows}, for the SCDA aperture, the difference in throughput between the initial state (blue dashed line) and final state (blue solid line) of the ACAD-OSM correction is $\Delta T_2$ = 3\% at 16 $\lambda_0/D_{ap}$. For the WFIRST aperture, the difference in throughput between the initial state (red dashed line) and final state (red solid line) of the ACAD-OSM correction is $\Delta T_3$ = 9\% at 16 $\lambda_0/D_{ap}$. 

Large struts have therefore not only an impact on the performance on contrast (for a given number of actuators) but also on the throughput of the correction, because of the high strokes they introduce on the DMs. The discontinuities due to the segmentation (only present in the SCDA aperture) have a less important impact on these metrics than the width of the struts. This is due to the the fact that segment discontinuities introduce less low-order spatial frequencies in the aperture, and therefore have a limited impact in the DH and can be corrected using less strokes on the DMs. However, segments in the aperture can fail, degrading the contrast in the DH. We study the correction of these segment failure in the next section.

%-------------------------------------------------------------------------------------------------
\begin{figure}
\begin{center}
 \includegraphics[width = .48\textwidth, trim= 0.1cm 2cm 4.5cm 2cm, clip = true]{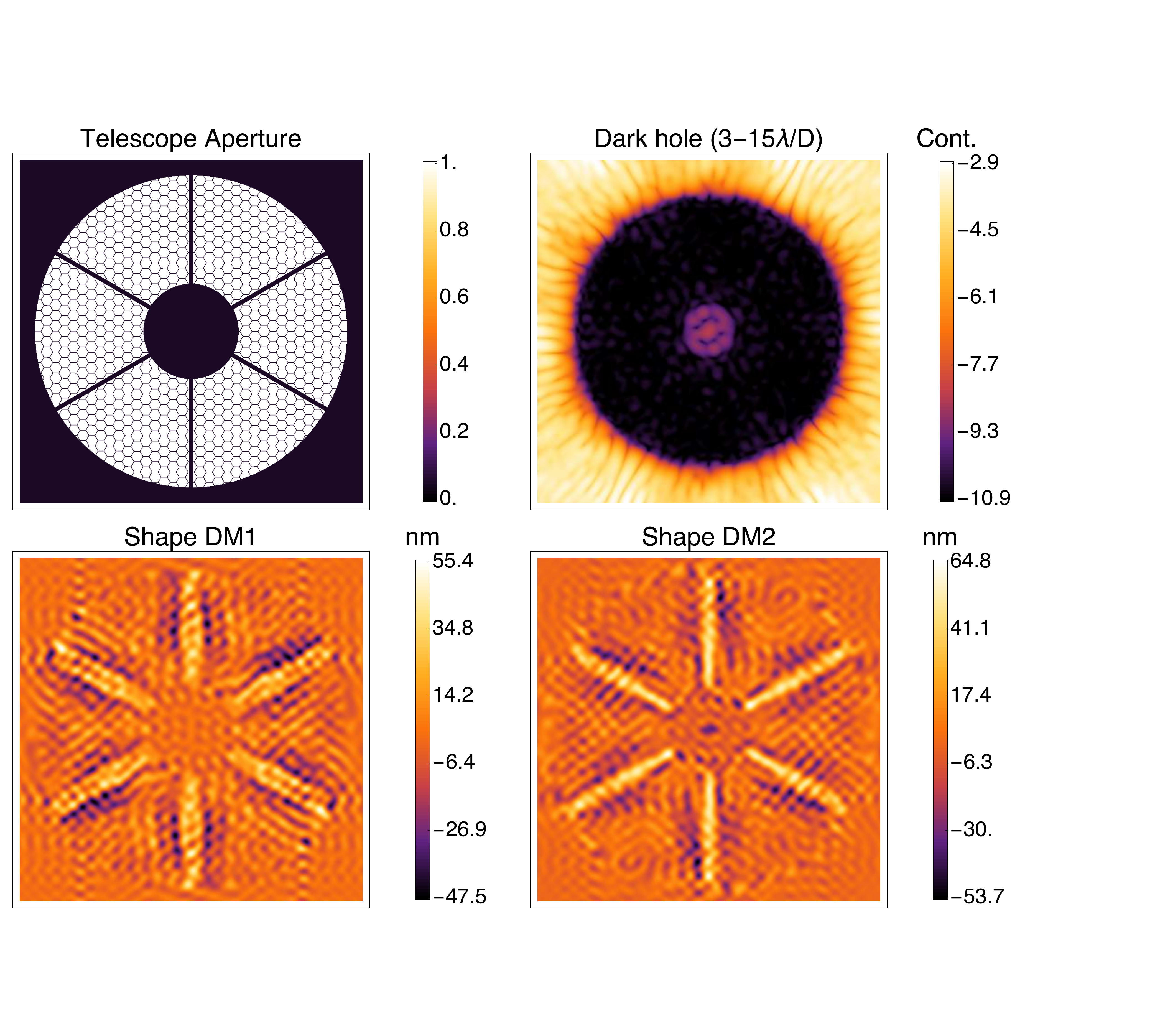}
 \end{center}
\caption[fig:eelt_dh]
{\label{fig:eelt_dh} E-ELT aperture (charge 6 PAVC, $N_{act} = 48$, IAP = 0.3 mm, D = $48 * 0.3$ mm, $z = 0.3$ m, $\Delta \lambda /\lambda_0 = $ 10\%). Top left: E-ELT aperture. Top right: the final 3-15 $\lambda_0/D_{ap}$ DH. Bottom: the DM shapes obtained using ACAD-OSM for this solution.}
\end{figure}
%-------------------------------------------------------------------------------------------------

%-------------------------------------------------------------------------------------------------
\subsection{Missing or inoperable aperture segments}
\label{sec:eelt}
%-------------------------------------------------------------------------------------------------

The aperture of the European Extremely Large Telescope (E-ELT, Figure~\ref{fig:eelt_dh}, top left) is characterized by a large central obscuration (30.5\% of the aperture outer radius), relatively large struts (4\% of the left area) to support the weight of the secondary mirror, and highly segmentation (hundreds of mirrors). With so many segments, a few of the segments may be inoperable or in maintenance during observations every day. Compared with techniques involving static mirrors or apodization, active methods can quickly adapt to any change in the aperture geometry. In this section, we study the capabilities of ACAD-OSM in the presence of missing or inoperable segments. The evolution of the aperture does not change the ACAD-OSM iterative process and the interaction matrix is built without prior knowledge of the missing segments.

\begin{figure}
\begin{center}
 \includegraphics[width = .48\textwidth, trim= 0.1cm 2cm 4.5cm 2cm, clip = true]{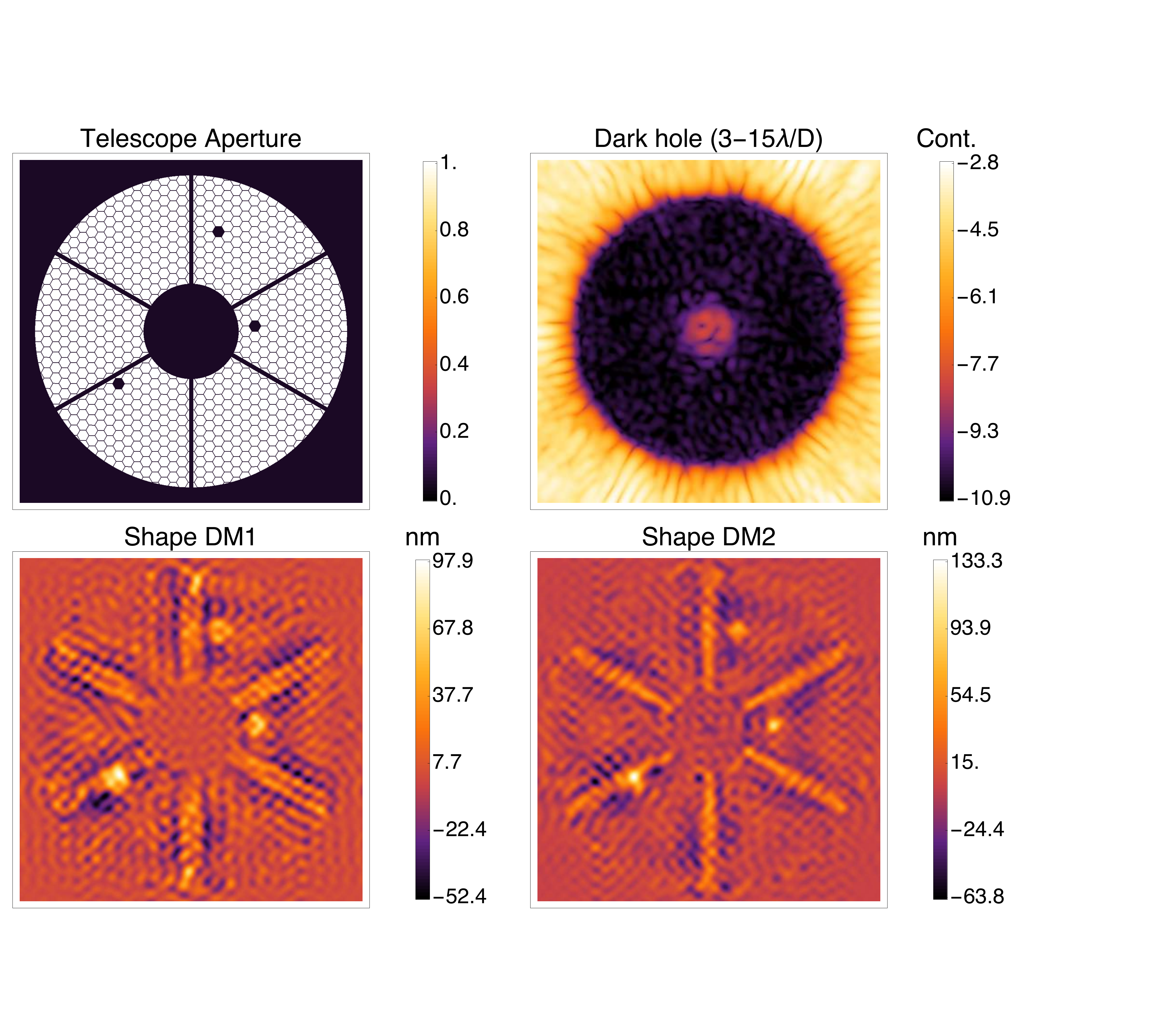}
 \end{center}
\caption[fig:eeltbs_dh]
{\label{fig:eeltbs_dh} E-ELT aperture with 3 missing segments (charge 6 PAVC, $N_{act} = 48$, IAP = 0.3 mm, D = $48 * 0.3$ mm, $z = 0.3$ m, $\Delta \lambda /\lambda_0 = $ 10\%). Top left: E-ELT aperture with 3 missing segments. Top right: the final 3-15 $\lambda_0/D_{ap}$ DH. Bottom: the DM shapes obtained using ACAD-OSM for this solution.}
\end{figure}

\begin{figure}
\begin{center}
 \includegraphics[trim= 1.8cm 0.8cm 1cm 0.5cm, clip = true, width = .48\textwidth]{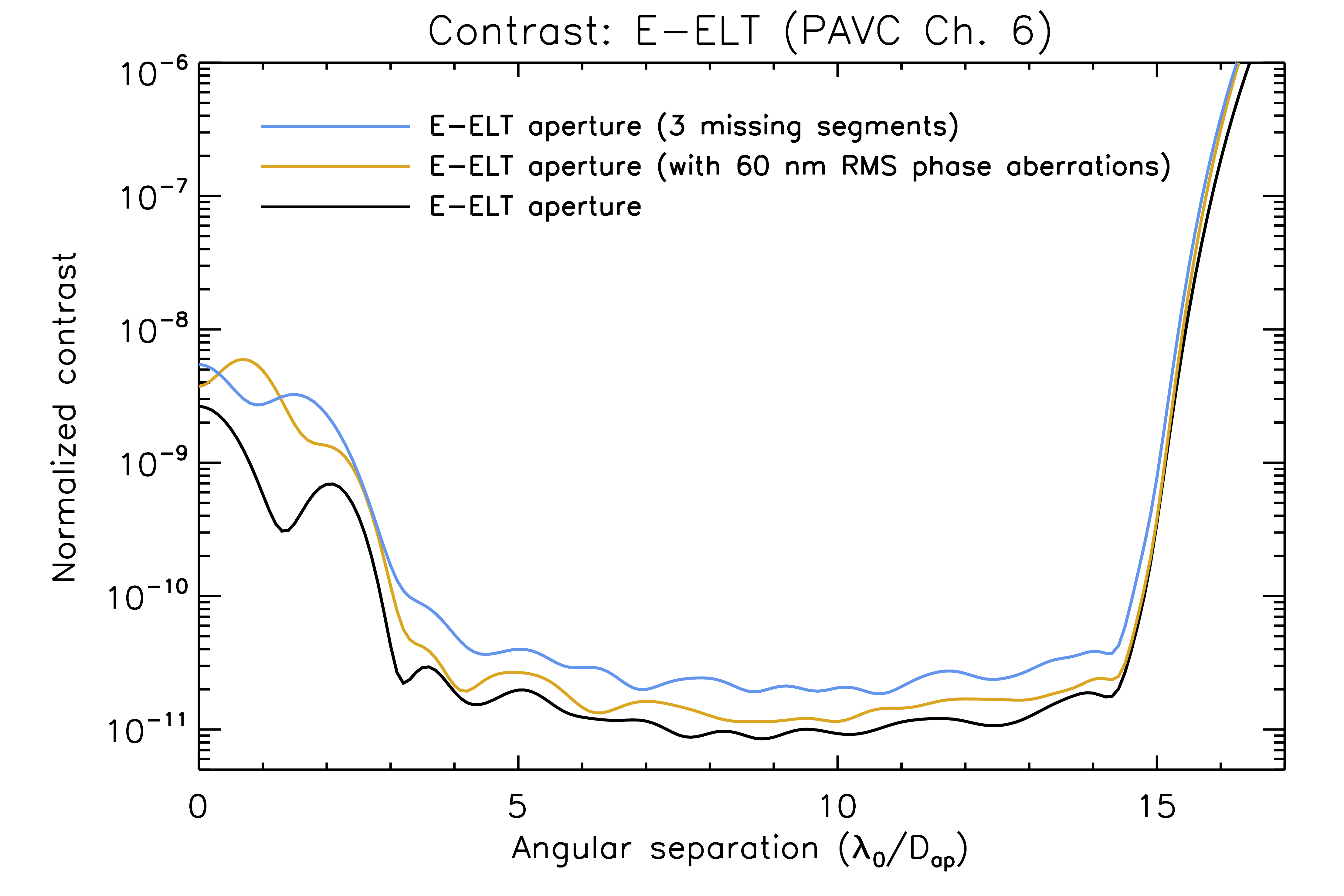}
  \end{center}
\caption[fig:eelt_contrast]
{\label{fig:eelt_contrast} Contrast level for the E-ELT aperture (charge 6 PAVC, $N_{act} = 48$, IAP = 0.3 mm, D = $48 * 0.3$ mm, $z = 0.3$ m, $\Delta \lambda /\lambda_0 = $ 10\% BW). Also shown, he contrast levels for the same aperture with 3 missing segments (blue solid line) and in presence of a 60 nm RMS phase error (yellow line).}
\end{figure}

The results for the highly segmented E-ELT aperture are shown in Fig.~\ref{fig:eelt_dh} in the absence of missing segment and in Fig~\ref{fig:eeltbs_dh} with three missing or inoperable segments, on a 3-15 $\lambda_0/D_{ap}$ DH. Fig.~\ref{fig:eelt_contrast} shows the contrast results in presence of missing segments (blue solid lines) in comparison with the nominal configuration with no missing segments (dark lines). The throughput is barely impacted by the 3 missing segments. ACAD-OSM compensates for the image degradation caused by the missing segments, reaching within a factor of 3 of the nominal contrast inside the 3-15 $\lambda_0/D_{ap}$ DH. 

\begin{figure}
\begin{center}
 \includegraphics[trim= 2.5cm 0.8cm 1cm 0.5cm, clip = true,width = .48\textwidth]{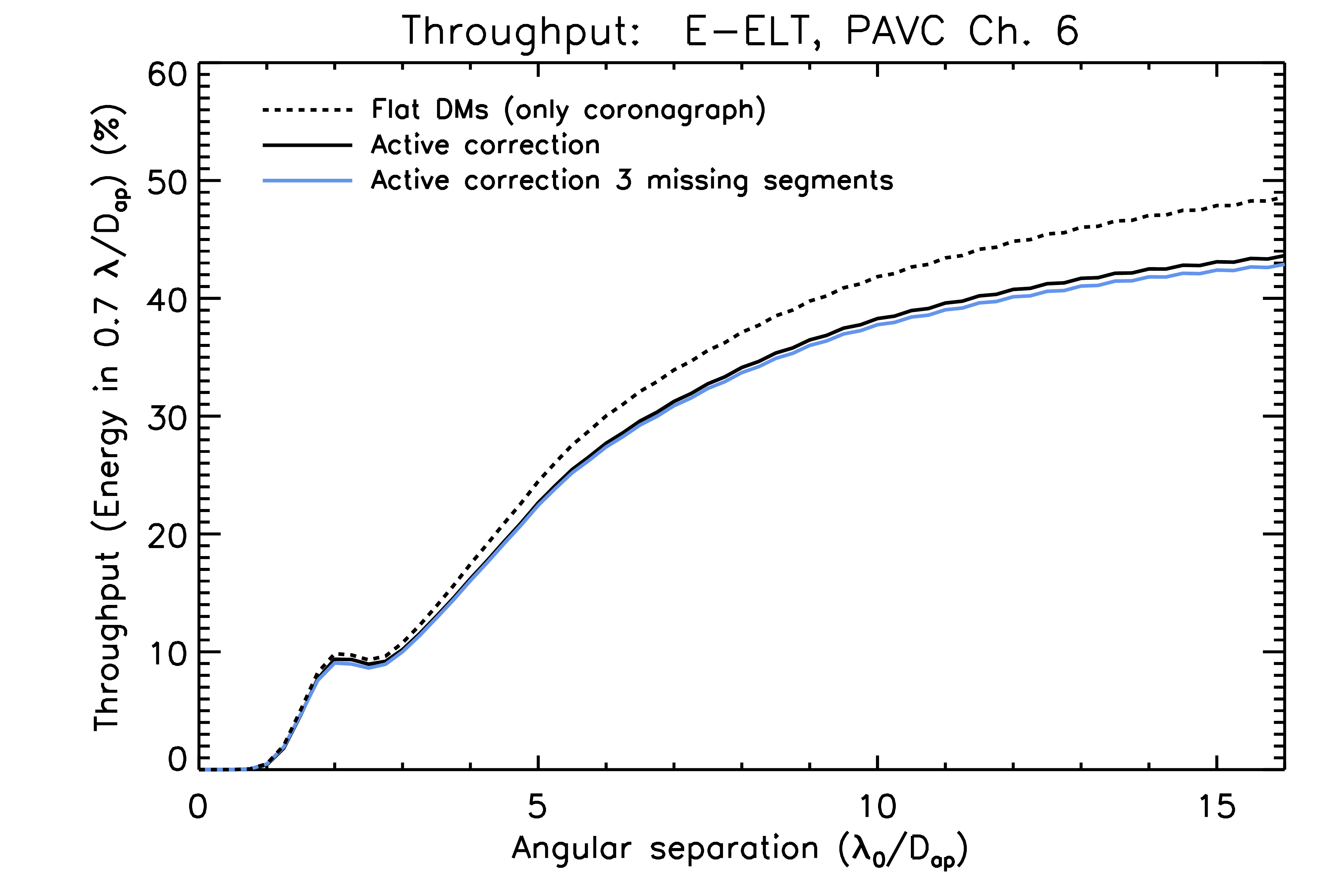}
  \end{center}
\caption[fig:eelt_throughput]
{\label{fig:eelt_throughput} Throughput level for the E-ELT aperture. The dashed line shows the throughput before any correction (due to the charge 6 PAVC alone). The solid line shows the throughput at the end of the  correction ($N_{act} = 48$, IAP = 0.3 mm, D = $48 * 0.3$ mm, $z = 0.3$ m, $\Delta \lambda /\lambda_0 = $ 10\% BW) for the E-ELT aperture (black line) and for the E-ELT aperture  with 3 missing segments (blue solid line).}
\end{figure}

Fig~\ref{fig:eelt_throughput} shows the impact on throughput of the correction of missing segments. The difference between the nominal aperture (black solid curve) and the missing segment aperture (blue solid curve) is inferior to the percent. 

%-------------------------------------------------------------------------------------------------
%-------------------------------------------------------------------------------------------------

%-------------------------------------------------------------------------------------------------
%-------------------------------------------------------------------------------------------------

%-------------------------------------------------------------------------------------------------
%-------------------------------------------------------------------------------------------------

In this section, we first showed that the ACAD-OSM technique is capable of reaching the $10^{-10}$ contrast limit for several apertures (WFIRST, SCDA, E-ELT), with a 10\% bandwidth. We also notice that for non-axisymmetric discontinuities, the width of the struts is the main driver of the performance in contrast and throughput. In the next section, we expand the size of the spectral BW and measure its influence on the results. 

%-------------------------------------------------------------------------------------------------
%-------------------------------------------------------------------------------------------------
%-------------------------------------------------------------------------------------------------
\section{Bandwidth} 
\label{sec:bandwidth}
%-------------------------------------------------------------------------------------------------
%-------------------------------------------------------------------------------------------------
%-------------------------------------------------------------------------------------------------

\begin{figure}
\begin{center}
 \includegraphics[width = .48\textwidth, trim= 0.1cm 4.5cm 4.5cm 4cm, clip = true]{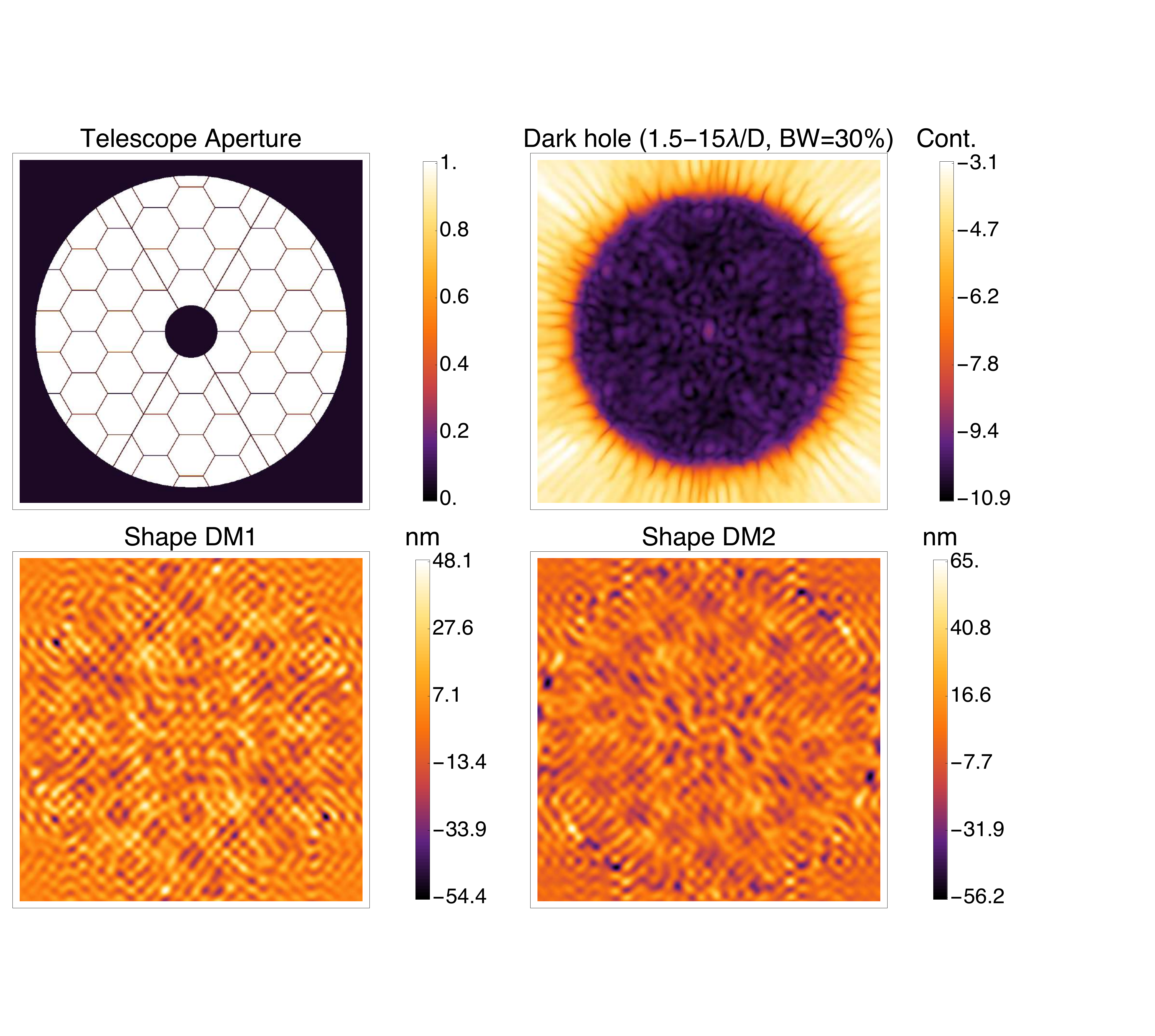}
 \end{center}
\caption[fig:scda_dh]
{\label{fig:scda_dh} SCDA aperture with a Charge 6 PAVC coronagraph, $N_{act} = 48$, IAP = 0.3 mm, D = $48 * 0.3$ mm, $z = 0.3$ m and a $\Delta \lambda /\lambda_0 = $ 30\% BW. Top left: SCDA aperture. Top right: the final 1.5-15 $\lambda_0/D_{ap}$ DH. Bottom: the DM shapes obtained using ACAD-OSM for this solution.}
\label{fig:scda_dh_ACADOSM}
\end{figure}

\begin{figure}
\begin{center}
 \includegraphics[trim=  1.8cm 0.8cm 1cm 0.5cm, clip = true,width = .48\textwidth]{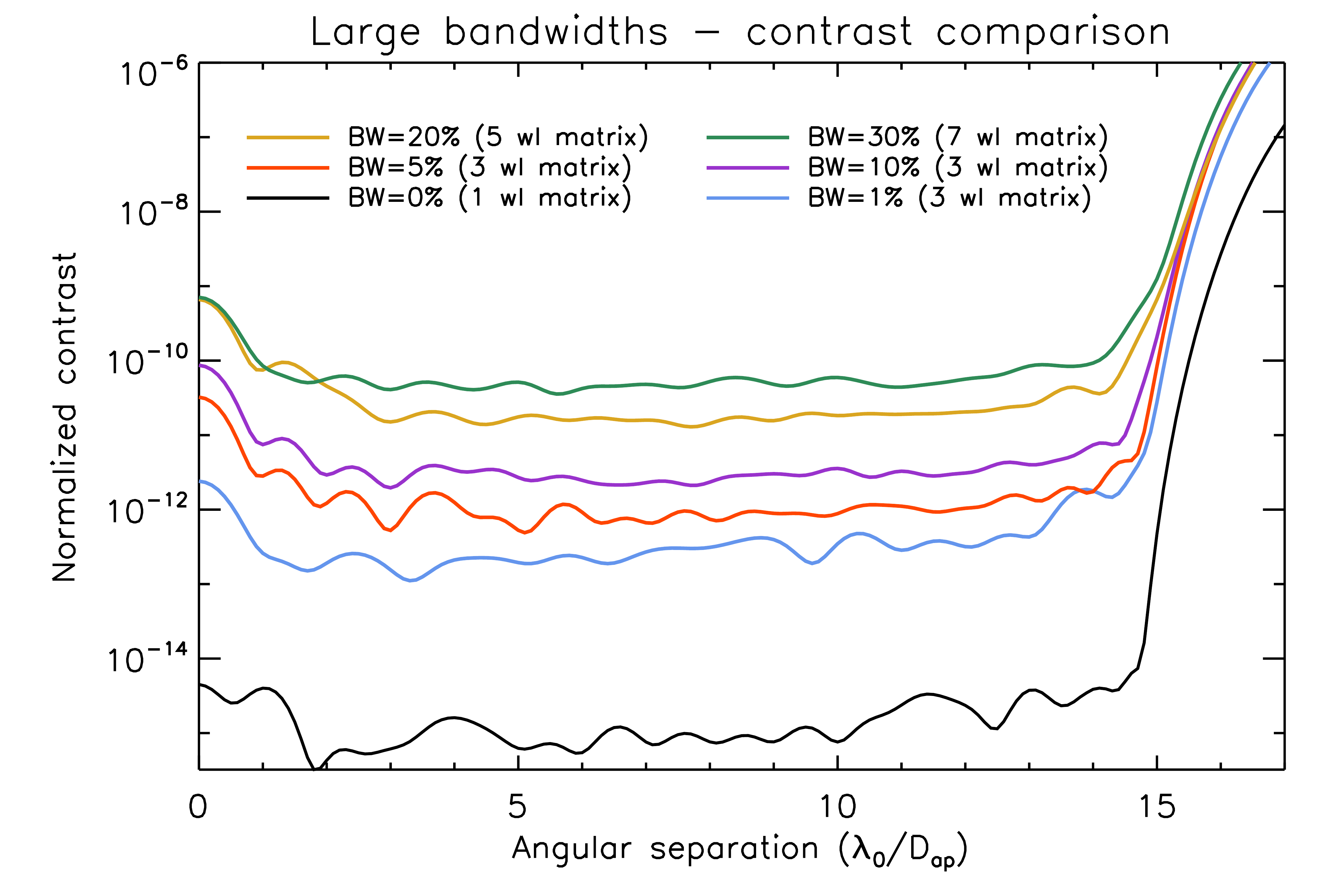}
  \end{center}
\caption[fig:bw_contrast]
{\label{fig:bw_contrast} Contrast levels for different BWs (around $\lambda_0 = 550$ nm) for the SCDA aperture, a charge 6 PAVC and $N_{act} = 48$, IAP = 0.3 mm, D = $48 * 0.3$ mm, $z = 0.3$ m for DM setup.}
\end{figure}

\begin{figure}
\begin{center}
 \includegraphics[trim= 2.5cm 0.8cm 1cm 0.5cm, clip = true,width = .48\textwidth]{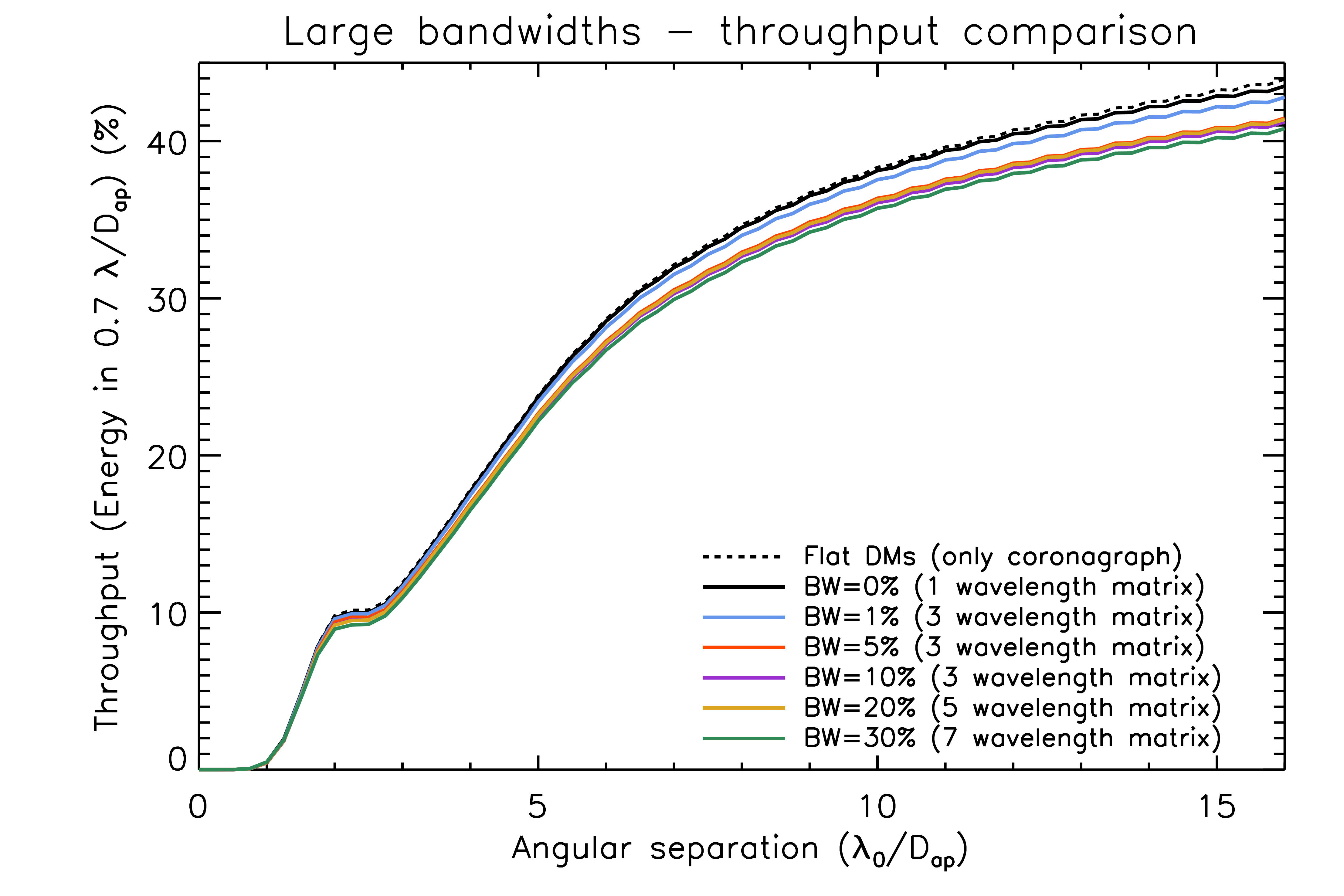}
  \end{center}
\caption[fig:bw_throughput]
{\label{fig:bw_throughput}Throughput levels for different BWs (around $\lambda_0 = 550$ nm) for the SCDA aperture, a charge 6 PAVC and $N_{act} = 48$, IAP = 0.3 mm, D = $48 * 0.3$ mm, $z = 0.3$ m for DM setup. The dashed line shows the throughput before any correction (due to the charge 6 PAVC alone).}
\end{figure}

Fig~\ref{fig:scda_dh_ACADOSM} shows the results with a charge 6 PAVC with the SCDA aperture and a larger bandwidth ($\Delta \lambda /\lambda_0 = $ 30\% BW), with $N_{act} = 48$, IAP = 0.3 mm, D = $48 * 0.3$ mm, $z = 0.3$ m. Contrast and throughput levels associated with this DH are shown with green curves in Fig.~\ref{fig:bw_contrast} and Fig.~\ref{fig:bw_throughput}.

%-------------------------------------------------------------------------------------------------
%-------------------------------------------------------------------------------------------------

To achieve a broadband correction, the interaction matrices are built by concatenating interaction matrices at different wavelengths that are centered around the central wavelength and equally spatially sampled one to another over the BW. We studied the influence of the number of sampling wavelengths on the performance of the correction. First, the number of sampling wavelengths has a linear impact on the size of the matrix and therefore on the calculation time. Secondly, it has a limited impact on the performance in contrast level: for a coarsely sampled correction (e.g. 3 wavelength matrix for a 30\% BW correction), the contrast inside the DH remains the same as for a finely sampled correction (e.g. 7 wavelength matrix for a 30\% BW correction), but the DH at the outer working angle (OWA) becomes less steep, practically decreasing the OWA slightly. Finally, the performance in throughput decreases negligibly when lowering the sampling.

\begin{figure}
\begin{center}
 \includegraphics[trim=  1.8cm 0.8cm 1cm 0.5cm, clip = true,width = .48\textwidth]{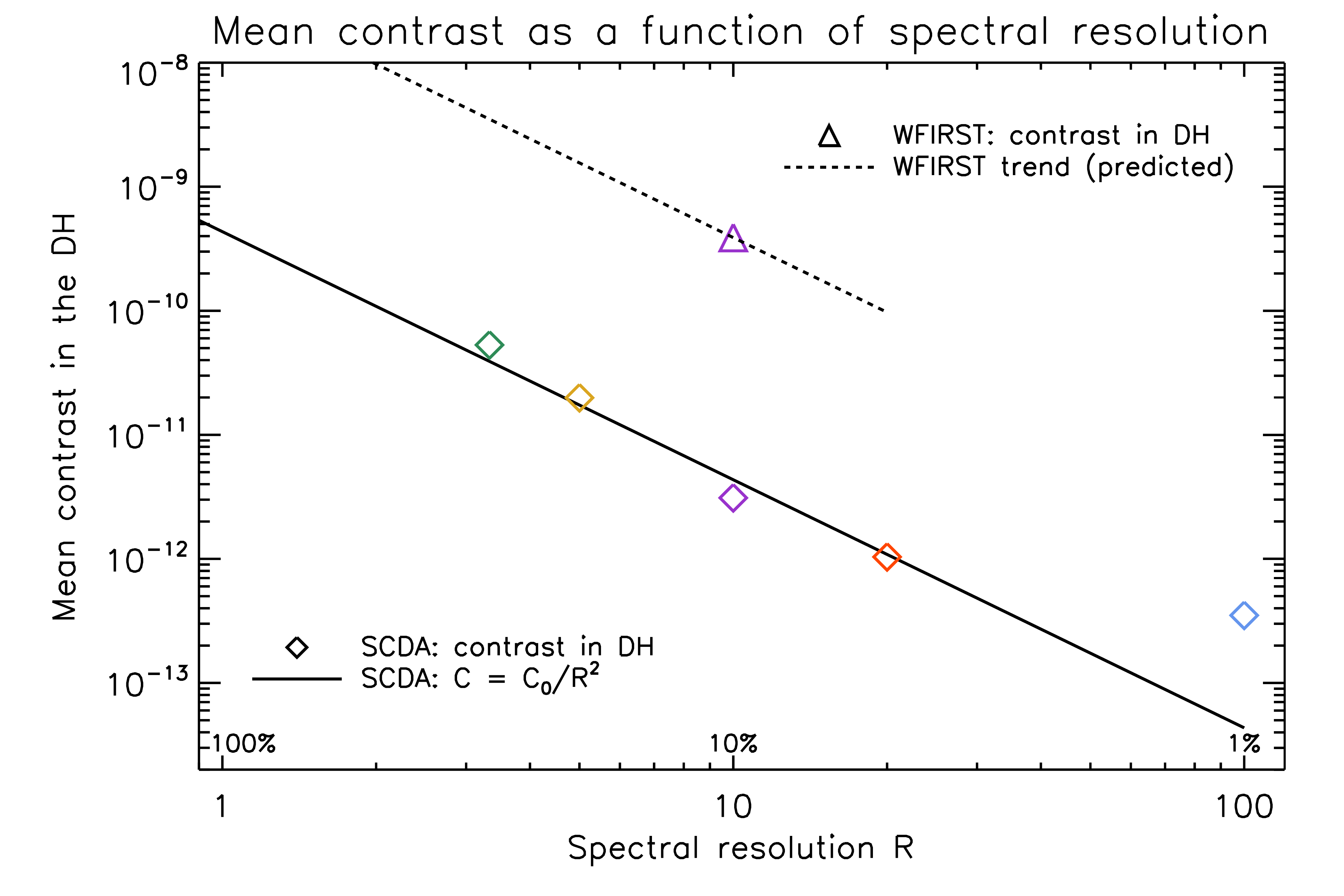}
  \end{center}
\caption[fig:contrast_function_R]
{\label{fig:contrast_function_R} Contrast levels as a function of spectral resolution R. The diamonds represent the results obtained with the SCDA aperture (charge 6 PAVC and $N_{act} = 48$, IAP = 0.3 mm, D = $48 * 0.3$ mm, $z = 0.3$ m), with colors matching Fig~\ref{fig:bw_contrast}. The solid line is the fit of these results using Eq.~\ref{eq:bw_shaklan06}. The triangle shows the contrast obtained with a 10\% bandwith (R = 10) for the WFIRST aperture (charge 6 PAVC, $N_{act} = 48$, IAP = 1 mm, D = $48*1$ mm, $z = 1$ m). The dashed line shows predicted contrast performance for this aperture.}
\end{figure}

Figure \ref{fig:bw_contrast} show the results in contrast for different widths of the spectral band (around $\lambda_0 = 550$ nm). The mean value of contrast level in the DH is plotted with diamonds in Fig.~\ref{fig:contrast_function_R}, as a function of the spectral resolution $R = \lambda/\Delta\lambda$. Indeed, \cite{shaklan06} predict that the degradation in contrast with the BW follows (Eq. 12):

\begin{equation}
\label{eq:bw_shaklan06}
C = \dfrac{C_0}{R^2}\,\,.
\end{equation}
where $C$ denotes the mean contrast for R = 1. $C_0$ is a constant, depending on the coronagraph, the aperture geometry, the DH size and the DM setup (see ACAD-OSM II). The contrast follows this trend for a BW between 5\% (R = 20) and 30\% (R = 3.3), see black solid line in Fig.~\ref{fig:contrast_function_R}, with $C_0 = 4.3\times10^{-10}$. The ratio between the mean contrasts that are measured and predicted by this curve is always smaller than 1.4. This trend does not apply to the 1\% BW (R= 100) results, 10 times worse than expected and for monochromatic light case, probably because other factors might limit the contrast at that point. However, using this curve, results with the ACAD-OSM method can be predicted for a larger BW (40 - 100\% or R = 2.5 - 1). Using only one point (one bandwidth performance in contrast), the performance of the technique can be predicted at any BWs. For example, the predicted contrast for large BW for the WFIRST aperture and DM setup is plotted with a dashed line ($C_0 = 3.9\times10^{-8}$), using the mean contrast in the 3-10 $\lambda_0/D_{ap}$ DH at BW = 10\% (DM setup $N_{act} = 48$, IAP = 1 mm, D = $48 * 1$ mm, $z = 1$ m, represented by a magenta triangle).

Finally, Fig.~\ref{fig:bw_throughput} shows the results in throughput for the same bandwidth. The dashed line shows the throughput before any correction, with flat DMs (due to the charge 6 PAVC alone). The colored solid curves show the throughput for several BWs. The throughput degrades as the BW increases but the loss remains limited to a few percents.

One has to be careful when using this law to predict the contrast at different bandwidth, which only apply for 2 DM setup in the Talbot-effect-limited regime, as defined in ACAD-OSM II.

%-------------------------------------------------------------------------------------------------
%-------------------------------------------------------------------------------------------------

%-------------------------------------------------------------------------------------------------
%-------------------------------------------------------------------------------------------------

%-------------------------------------------------------------------------------------------------
%-------------------------------------------------------------------------------------------------

%-------------------------------------------------------------------------------------------------
\section{Correction of misalignments in the coronagraph design}
\label{sec:misalignments}
%-------------------------------------------------------------------------------------------------

Section~\ref{sec:compar_fix_apodiz} showed that using a combination of static apodization and active correction gives comparable results as a standalone apodization optimized for an aperture with all its discontinuities. However, a static apodization only accounts for known aberrations or discontinuities in the optical design. In this paper, we show the advantages of using an active and iterative method that only relies on an interaction matrix built without prior knowledge the state of the system (aperture, alignment, aberrations, or even type of coronagraph). In Section~\ref{sec:eelt}, we proved that ACAD-OSM is robust to changes in the aperture. In this section, we study the impact of misalignments in the coronagraph system.

\begin{figure}
\begin{center}
 \includegraphics[trim= 1.5cm 1cm 1.0cm 0.5cm, clip = true,width = .48\textwidth]{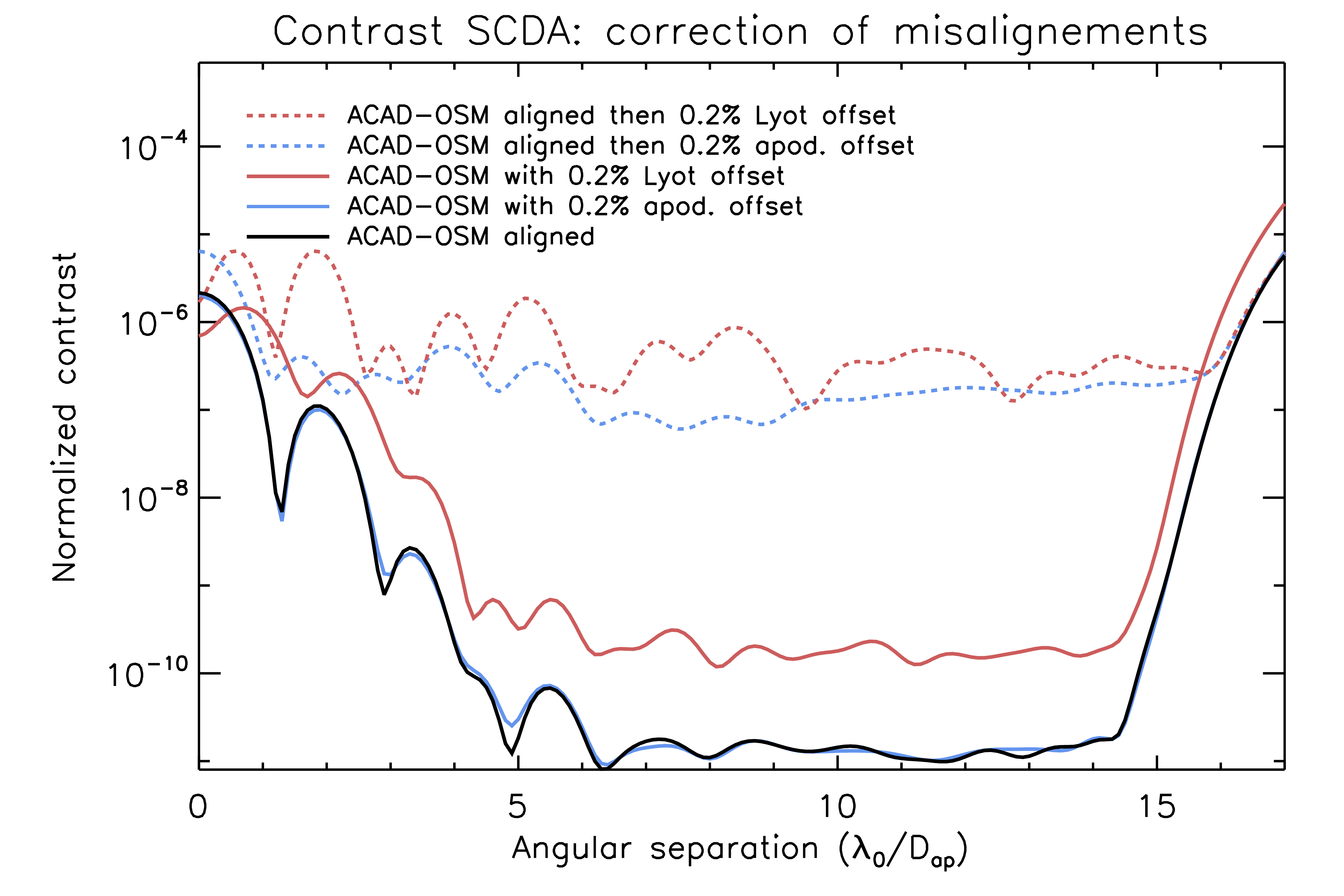}
  \end{center}
\caption[fig:contrast_scda_aplc_compar_misal]
{\label{fig:contrast_scda_aplc_compar_misal} Contrast results for the SCDA aperture  with an APLC with and without mi-alignments. In black solid line is shown the results for an aligned APLC with an apodizer that is only optimized for the central obscuration in addition to a ACAD-OSM system. In yellow, the same system is used, but with a misalignment of 0.2\% in diameter of the apodization, corrected by the ACAD-OSM algorithm. In red, the same system is used, but with a misalignment of 0.2\% in diameter of the Lyot stop, corrected by the ACAD-OSM algorithm. The DM setup is $N_{act} = 48$, IAP = 0.3 mm, D = $48 * 0.3$ mm, $z = 0.3$ m, and the BW is $\Delta \lambda /\lambda_0 = $ 10\%.}
\end{figure}

Several types of misalignment between the optics can degrade the contrast, e.g. a misalignment of the apodizer with respect to the re-imaged aperture in the entrance pupil plane. This problem is usually fixed by designing an apodizer for an aperture with oversized central obstruction and discontinuities, but such a solution leads to a relative loss in transmission. Optimizing an apodization for a centrally obstructed pupil only solves the issue for errors in rotation along the optical axis but leaves the problem for translation errors. Here, we analyze the response of ACAD-OSM to a small misalignment in translation of the apodization in the system. For this test, an APLC with an apodizer optimized for a centrally obstructed pupil is used and misaligned by a 0.2\% $D_{ap}$ with respect to the optical axis. The results in Figure~\ref{fig:contrast_scda_aplc_compar_misal} shows in black the initial results in contrast, after ACAD-OSM, when the system is aligned. The dashed blue line shows the degradation of contrast of this ACAD-OSM correction when a small misalignment of the apodization is introduced. 

\begin{figure}
\begin{center}
 \includegraphics[trim= 2.5cm 1cm 1.0cm 0.5cm, clip = true,width = .48\textwidth]{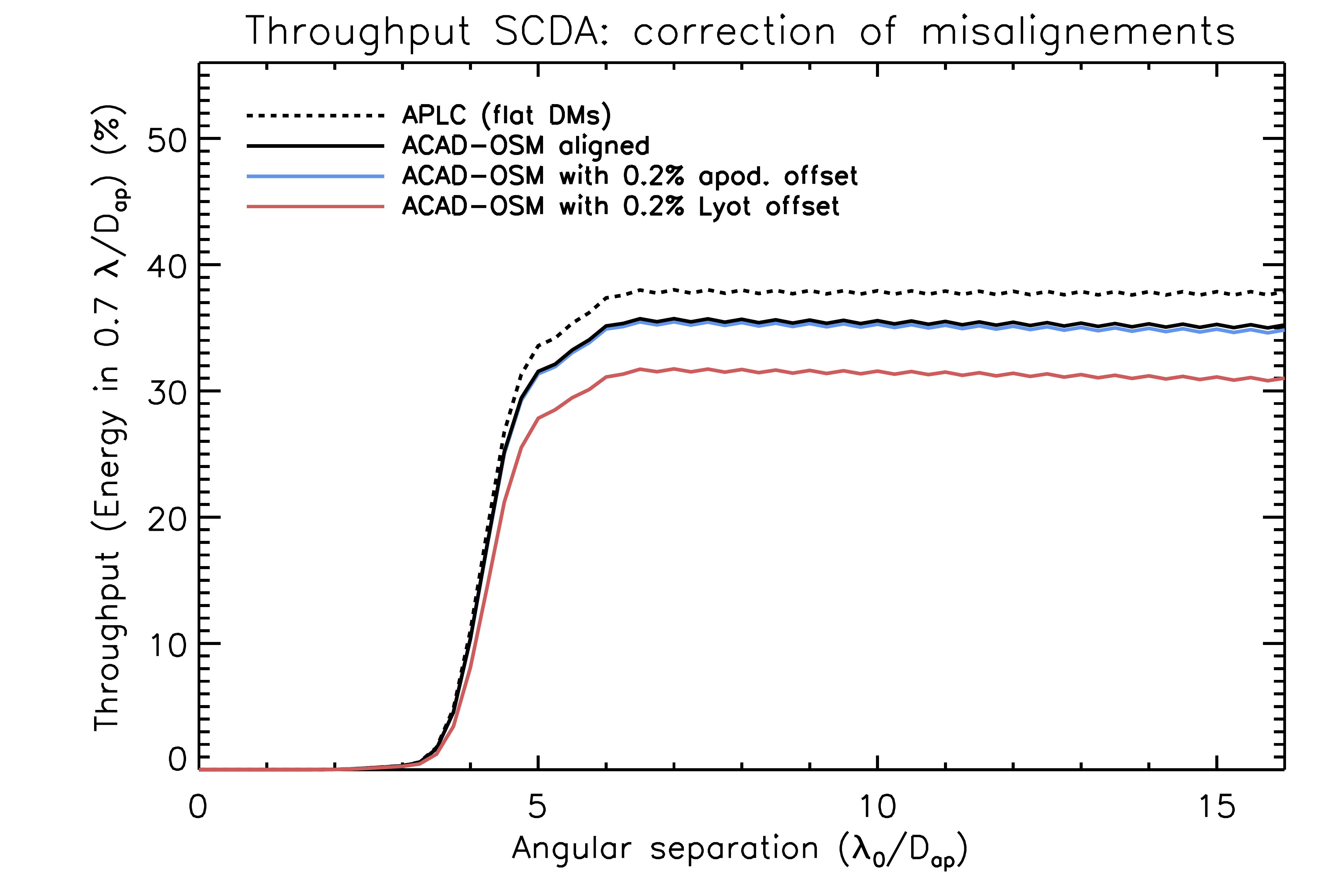}
  \end{center}
\caption[fig:throughput_scda_aplc_compar_misal]
{\label{fig:throughput_scda_aplc_compar_misal} Throughput results for the SCDA aperture with an APLC with and without mi-alignments. The dashed line shows the throughput with flat DMs, before any correction (due to the APLC alone). In black solid line is shown the results for an aligned APLC with an apodizer that is only optimized for the central obscuration in addition to a ACAD-OSM system. In yellow, the same system is used, but with a misalignment of 0.2\% in diameter of the apodization, corrected by the ACAD-OSM algorithm. In red, the same system is used, but with a misalignment of 0.2\% in diameter of the Lyot stop, corrected by the ACAD-OSM algorithm. The DM setup is $N_{act} = 48$, IAP = 0.3 mm, D = $48 * 0.3$ mm, $z = 0.3$ m, and the BW is $\Delta \lambda /\lambda_0 = $ 10\%.}
\end{figure}

We try to apply the ACAD-OSM system, this time in presence of an apodization misalignment. Once again, the evolution of the system does not change the ACAD-OSM iterative process and the interaction matrix is built without prior knowledge of this misalignment.Starting the correction from the initial state (flat DMs) with this misalignment, the results obtained (blue solid curve) are similar to the ones obtained in the aligned case. Fig.~\ref{fig:throughput_scda_aplc_compar_misal} shows the results in throughput. The dash black line is the throughput with flat DMs (initial state). The result in the aligned case (black solid line) and if the apodization is misaligned (blue solid line) are almost identical. This algorithm is therefore robust to a small misalignment of the apodizer in the pupil plane.

Finally, some apodized coronagraph designs can suffer from a misalignment of the Lyot stop with respect to the entrance pupil of the coronagraph. Two sorts of strategies have been developed for the design of the apodization. Some apodizations simply cancel out the energy from an on-axis star image inside the Lyot stop \citep[e.g.][]{soummer03,mawet13,fogarty17}. Their corresponding coronagraphs are not sensitive to Lyot stop misalignments and a simple undersizing of the outer radius and oversizing of the inner radius can prevent this problem. However, other apodizations coherently recombines the electric field in the Lyot plane to produce a DH in the final image plane \citep{kasdin03,ruane15,ruane16,ndiaye15b, ndiaye16, zimmerman16, zimmerman16spie}. Their corresponding coronagraphs are however very sensitive to a Lyot stop misalignment which breaks the fine recombination of coherent light in the relayed pupil plane and leads to a contrast degradation in the final image plane. The response of ACAD-OSM to this effect is studied with the APLC design where the Lyot stop is decentered by a 0.2\%$D_{ap}$ with respect to the optical axis. The effect of this misalignment are shown in Figure~\ref{fig:contrast_scda_aplc_compar_misal}. The correction is first made with a perfectly aligned system (black solid line) then a misalignment of the Lyot is introduced, degrading the contrast (dashed red line). Starting the correction from the initial state (flat DMs) with this Lyot misalignment, the results obtained (red solid curve) shows that ACAD-OSM can partially compensate for the loss in contrast. Fig.~\ref{fig:comparAPLC_throughput} (red solid curve) shows that this correction is done at the expense of some throughput. Finally, we also tested (successfully, but not shown in this paper) to use ACAD-OSM to correct for a Lyot misalignment introduced after a full aperture-opitimized APLC, like the one presented in Section~\ref{sec:compar_fix_apodiz}.

With hundreds of pixels in each of the apodizer directions, static apodizations control higher spatial frequencies and achieve larger and often deeper DH than active correction algorithms. However, the flexibility of an active method proves relevant to the evolution of the system with time, a situation for which static systems are not very robust. Static apodizers and active techniques can benefit from each other to achieve better and more robust performance. 

%-------------------------------------------------------------------------------------------------
%-------------------------------------------------------------------------------------------------

%-------------------------------------------------------------------------------------------------
%-------------------------------------------------------------------------------------------------

%-------------------------------------------------------------------------------------------------
\section{Phase errors}
\label{sec:phase_errors}
%-------------------------------------------------------------------------------------------------

Finally, this technique can be used without changes to correct for small phase and amplitude errors. In this section are shown the capabilities of this technique in the presence of both aperture discontinuities and realistic phase errors with for example low order aberrations caused by segments misalignments (Sec.~\ref{sec:phase_errors_segments}) or with higher order caused by optical aberrations (Sec.~\ref{sec:high_order_phase_errors}).

%-------------------------------------------------------------------------------------------------
\subsection{Influence of small phase errors due to segment misalignments}
\label{sec:phase_errors_segments}
%-------------------------------------------------------------------------------------------------

\begin{figure}
\begin{center}
 \includegraphics[width = .48\textwidth, trim= 0.1cm 4.5cm 4.5cm 4cm, clip = true]{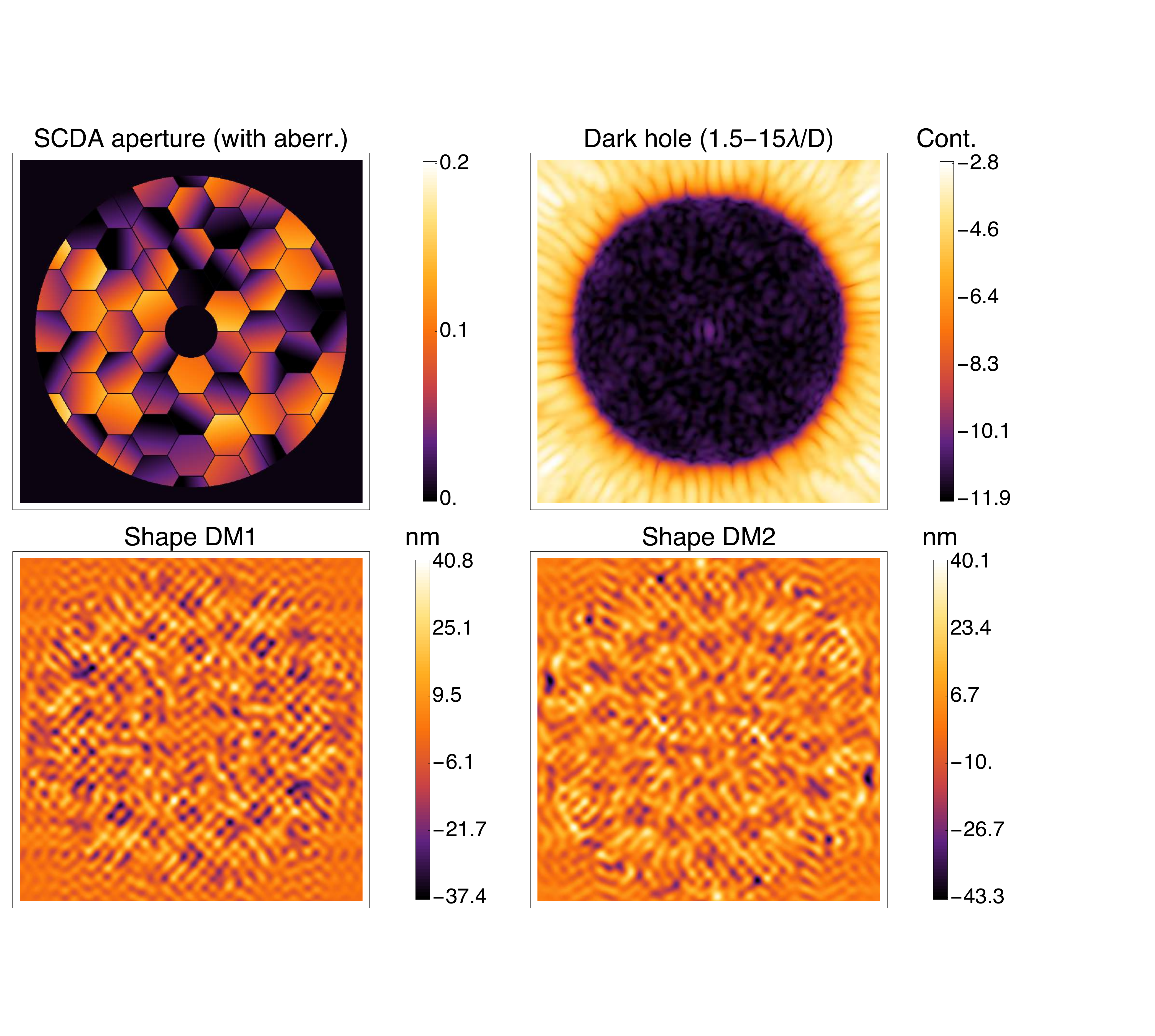}
 \end{center}
\caption[fig:scda_dh_segmentphase]
{\label{fig:scda_dh_segmentphase} SCDA aperture (Charge 6 PAVC coronagraph, $N_{act} = 48$, IAP = 0.3 mm, D = $48 * 0.3$ mm, $z = 0.3$ m) with a 10\% BW in the presence of phase errors due to segment misalignments. Top left: SCDA aperture with phase aberrations. Top right: the final 1.5-15 $\lambda_0/D_{ap}$ DH. Bottom: the DM shapes obtained using ACAD-OSM for this solution.}
\end{figure}

So far, all the examples have assumed a perfect cophasing of all the segments and therefore, the aperture only includes discontinuities, but no phase and amplitude aberrations. Several methods have been developed to align the segments of the primary to a low level of aberrations. A small amount of residual phase due to imperfect alignment of the segments can limit the contrast in the DH of the coronagraph, but can be corrected easily with ACAD-OSM. Figure~\ref{fig:scda_dh_segmentphase} shows a DH with this setup, in the presence of phase errors due to segment misalignments in piston (10 nm peak-to-valley) and tip-tilt (10 nm peak-to-valley), after the correction with ACAD-OSM. The contrast and throughput performance (not shown here) are almost identical with or wihtout these small phase aberrations.

%-------------------------------------------------------------------------------------------------
%-------------------------------------------------------------------------------------------------

%-------------------------------------------------------------------------------------------------
\subsection{Influence of high-order spatial frequency phase errors}
\label{sec:high_order_phase_errors}
%-------------------------------------------------------------------------------------------------

Since ACAD-OSM uses the focal plane electric field to drive the correction, it addresses aperture discontinuity effects in the focal plane exactly like the effects of continuous phase or amplitude apertures. However, one can wonder if the correction can be limited in the presence of important phase aberrations creating speckles, partially masking the effect of the aperture discontinuities. To study this phenomenon, a large phase error in the entrance pupil plane with the E-ELT aperture is introduced and the ACAD-OSM technique is used to attempt to simultaneously correct for both this phase and the aperture discontinuities simultaneously. The phase error is static and we leave the analysis of the active correction with dynamic phase error to a further study. The phase error has an amplitude of 60 nm RMS, with a power spectral density law of $f^{-3}$, where $f$ denotes the spatial frequency. Once again, no change have been made to the algorithm, and the interaction matrix is built without prior knowledge of the phase aberration.  The results in contrast are given in Fig.~\ref{fig:eelt_contrast} (yellow curve), showing no significant difference with the case of the standalone E-ELT aperture. The observed throughput levels are similar (not shown here). The total number of iterations is also very similar (1157 and 1142 for the correction with and without phase aberration).

In this section, we showed that although the ACAD-OSM technique have been invented for the correction of aperture discontinuities, it can also simultaneously correct for phase aberrations introduced in the aperture, even in the case when these aberrations are partially masking the effect of the aperture discontinuities. A more complete study specifically on the correction of phase aberrations with a two DM coronagraphic design can be found in \cite{beaulieu17}.

%-------------------------------------------------------------------------------------------------
%-------------------------------------------------------------------------------------------------
%-------------------------------------------------------------------------------------------------
\section{Conclusion}
%-------------------------------------------------------------------------------------------------
%-------------------------------------------------------------------------------------------------
%-------------------------------------------------------------------------------------------------

In this paper, we presented the active correction of aperture discontinuities - optimized stroke minimization, ACAD-OSM, a new active method to correct for the discontinuities in the aperture. This method uses an adaptive interaction matrix to control two deformable mirrors. In the first part, we described the algorithm and its superiority over the ACAD technique introduced by \cite{pueyo13}. We have illustrated its capabilities with some specific cases (WFIRST, E-ELT, and an example of a realistic segmented space aperture). As the main results of this paper, (1) the ACAD-OSM method applies to any coronagraph or aperture and reaches $10^{-10}$ contrast level with a 10\% bandwidth and existing DM configurations; (2) it achieves high contrast at large bandwidths (test at bandwidths up to 30\% have been performed in this paper), following the theoretical limit predicted by \cite{shaklan06}; (3) this active technique  handles --- without any revision to the algorithm --- evolving or unknown optical aberrations or discontinuities in the pupil, including optical design misalignments, missing segments and/or phase errors.

In conclusion, a static optical design (apodization or static mirrors) alone is not the preferable approach to compensate for aperture discontinuities. We instead advocate for an active technique with two-DM correction, due to its adaptability to evolving or unknown optical design, such as coronagraph optic misalignments or static phase errors. However, we acknowledge that the current generation of DMs does not offer enough degrees of freedom to correct by itself for all the features of a complex aperture. In particular, the correction of the central obscuration with DMs only, is for the moment out of reach since it involves very large strokes. A solution combining the advantage of both methods (static and active correction), such as the one presented in this paper, should be pursued for future coronagraphic instruments.

Several parameters have a strong impact on the results in contrast and throughput that are obtained with this method. In particular, the design of the coronagraph and the DM setup (size of the DMs, number of actuators and distance between the DMs) can be optimized to achieve the best performance. The optimization of this technique for future large space mission by analyzing the effects and optimization of these parameters on its performance (in contrast, throughput, and robustness to jitter aberrations) will be done in the ACAD-OSM II paper.

\acknowledgments{JM is currently funded by the NSF. This material is based upon work carried out under subcontract \#1496556 with the Jet Propulsion Laboratory funded by NASA and administered by the California Institute of Technology. The authors would like to thank John Krist (JPL), Pierre Baudoz (LESIA) and Sylvain Egron (STScI/ONERA/LAM) for valuable discussions. We are grateful to the referee for his very constructive inputs that have greatly improved the overall presentation and clarity of this paper.}

\bibliographystyle{apj} 
\bibliography{acad_bib}  

\end{document}